\documentclass[11pt]{article}
\usepackage{geometry} 
\usepackage[english]{babel}
\usepackage{graphicx}

\begin{document}

\begin{center}
\textbf{Deuteron: properties and analytical forms of wave function
in coordinate space}
\end{center}

\begin{center}
\textbf{V. I. Zhaba}
\end{center}

\begin{center}
\textit{Uzhgorod National University, Department of Theoretical
Physics,}
\end{center}

\begin{center}
\textit{54, Voloshyna St., Uzhgorod, UA-88000, Ukraine}
\end{center}

\begin{center}
\textit{(Received June 26, 2017)}
\end{center}



\textbf{Abstract}

Static parameters of the deuteron, obtained by the wave functions
for various potential models, have been chronologically
systematized. The presence or absence of knots near the origin of
coordinates for the radial wave function of the deuteron have been
shown. Analytical forms for the deuteron wave function in
coordinate space have been reviewed. Both analytical forms and
parameterizations of the deuteron wave function, which are
necessary for further calculations of the characteristics of the
processes involving the deuteron, have been provided. In addition,
the asymptotic behaviors of deuteron wave function near the origin of coordinates and
for large values of distance have been analyzed in the paper.
Minimization of the number of numerically calculated coefficients
for new analytical forms as a product of exponential function
$r^{n}$ by the sum of the exponential terms
$A_{i}$\textit{*exp(-a}$_{i}*r^{3})$ have been done. The optimum
is $N$=7-10.

\textbf{1. Introduction}

Deuteron is the most elementary nucleus. He consists of the two
strongly interacting elementary particles: a proton and a neutron.
The simplicity and evidentness of the deuteron's structure makes
it a convenient laboratory for studying and modeling
nucleon-nucleon forces. Now, deuteron has been well investigated
both experimentally and theoretically.

The experimentally determined values of static properties of the
deuteron are in very much good agreement with the experimental
data. Owever despite that, there still are some theoretical
inconsistencies and problems. For example, in latest papers one
(for OBE \cite{Buck1979},  Bonn \cite{Machleidt2001} potentials)
or both (for Soft core Reid68 \cite{Reid1968},  Moscow
 \cite{Kukulin1998}, renormalized OPE and TPE chiral \cite{Arriola2007} potentials)
 components of the radial wave
function in coordinate space have knots near the origin of the
coordinates. The existence of knots in the wave functions of the
basic and sole state of the deuteron is the evidence of
inconsistencies and inaccuracies in implementation of numerical
algorithms in solving similar problems. Or it is connected with
features of potential models for the description of a deuteron.
The way the choice of numerical algorithms influences the solution
is shown in Refs. \cite{Haysak1,Haysak2,Bokhinyuk}. The knots of
the wave function in coordinate representation are analyzed in
more detail in the following sections of the article.

Besides, it should be noted that the deuteron wave function in
momentum space in the scientific literature is presented
ambiguously. In particular, in the S- component
\cite{Fujiwara2001,Garcon2001,Veerasamy2011,Fukukawa2015} (or in
S- and D- components \cite{Loiseau1987,Gross2010}), there is an
excess knot in the middle of interval for values of momentum.

It should also be noted that such potentials of the
nucleon-nucleon interaction as Bonn \cite{Machleidt2001}, Moscow
\cite{Kukulin1998}), Nijmegen group potentials (NijmI, NijmII,
Nijm93 \cite{stoks1994,Swart1995}), Argonne v18
\cite{Wiringa1995}, Paris \cite{Lacombe1980}, NLO, NNLO and N3LO
\cite{Epelbaum2005}, Idaho N3LO \cite{Epelbaum2015} or Oxford
potential \cite{downum2010} have quite a complicated structure and
cumbersome representation. Example, the original potential Reid68
was parameterized on the basis of the phase analysis by Nijmegen
group and was called as updated regularized version - Reid93. The
parametrization was done for 50 parameters of the potential, where
value \textit{$\chi $}$^{2}/N_{data}$=1.03
\cite{stoks1994,Swart1995}.

Besides, the deuteron wave function (DWF) in coordinate space can
be presented as a table: through respective arrays of values of
radial wave functions. It is sometimes quite difficult and
inconvenient to operate with such arrays of numbers during
numerical calculations. And the program code for numerical
calculations is bulky, overloaded and unreadable. Therefore, it is
feasible to obtain simpler and comfortabler analytical forms of
DWF representation. It is further possible on the basis to
calculate the form factors and tensor polarization, characterizing
the deuteron structure.

DWFs in a convenient form are necessary for use in calculations of
polarization characteristics of the deuteron, as well as to
evaluate the theoretical values of spin observables in \textit{dp}
scattering \cite{Ladygin1997}.

In addition to introduction, the first section and conclusions,
the article is composed of six more sections. The second section
deals with the deuteron wave function: main peculiarities and
scientific interest in its studying. The third section describes
the basic properties of the deuteron. The numerical values of
theoretical calculation results and experimental data are
presented in convenient tables. The fourth and the fifth sections
provide a description of basic analytical forms of DWF in the
coordinate representation. The sixth section describes the
"improved" analytical forms of DWF. The seventh section suggests
new analytical forms of DWF used in modern scientific literature.
Coefficients for new analytical forms in the form
$r^{n}$*$A_{i}$\textit{*exp(-a}$_{i}*r^{3})$ have been calculated.

The main objectives of the research in this paper are to
systematize the analytical forms of DWF in the coordinate
representation, calculate and analyze the coefficients for new
analytical forms.

\textbf{2. Deuteron wave function}

Wave function describes quant-mechanical system and is the basic
characteristic of microobjects. Knowledge of deuteron wave function allows
receiving the maximal information on system and theoretically to calculate
the characteristics measured on experiment. Deuteron wave function find as
the decision of system of coupled Schrodinger equations.

Deuteron wave functions write down as the sum of wave functions
for $^{3}$S$_{1}$- and $^{3}$D$_{1}$- state \cite{Blatt1958}

\begin{equation}
\label{eq1}
\Psi _d = \psi _S + \psi _D = \frac{u(r)}{r}Y_{101}^1 +
\frac{w(r)}{r}Y_{121}^1 ,
\end{equation}

where $u(r)$ and $w(r)$ are radial deuteron wave functions for
states with the orbital moments $l$=0 and 2; $Y_{JLS}^M (\theta
,\phi )$ are spherical harmonics, that are determined by orbital
moment $L$, spin $S$, the full moment $J=L+S$ and his projection
$M$ to an axis $z$. For deuteron: $S$=1; $J=M=S$=1.

The condition of normalization for DWF $\Psi _d $ can be written down as

\[
p_S + p_D = \int\limits_0^\infty {\left( {u^2(r) + w^2(r)} \right)dr} = 1,
\]

where $p_{S}$ and $p_{D}$ are probabilities to find out deuteron
in S- and D- state accordingly.

Taking into account spherical harmonics, it is possible to write down system
of the coupled differential equations of the second order for deuteron

\begin{equation}
\label{eq2}
\left\{ {\begin{array}{l}
 \frac{d^2u}{dr^2} + \left( { - k^2 - U_1 } \right)u = \sqrt {8 \cdot } U_T
w, \\
 \frac{d^2w}{dr^2} + \left( { - k^2 - \frac{6}{r^2} - U_2 } \right)w = \sqrt
8 U_T u. \\
 \end{array}} \right.
\end{equation}

Here $U_1 $, $U_2 $ are normalized potentials of channels $l=0; 2;
\quad U_3 $ are tensor component NN- interaction; $U_i (r) =
\frac{2\mu }{\hbar ^2}V_i (r)$; $k^2 = \frac{2\mu}{\hbar ^2}E$ is
wave number.

About the beginning of coordinates wave function D- state $w(r)$
has small value, because the repellent centrifugal barrier
$\frac{\hbar ^2l(l + 1)}{mr^2}$ will prevail on small distances.
Outside of radius for action of forces the behaviour for $w(r)$
also is determined by this barrier which sets asymptotic as
\cite{Brown1979}:

\[
w(r) \sim C\exp \left( { - \gamma r} \right)\left[ {1 + \frac{3}{\gamma r} +
\frac{3}{\left( {\gamma r} \right)^2}} \right].
\]

In paper \cite{naghdi2014} it was specified that one can divide
the main models into four categories: 1) the models based on
quantum chromo dynamics; 2) the effective field theory is another
outstanding approach to NN problem; 3) the boson exchange models;
4) the almost pure phenomenological NN potentials. Last decades
the second and fourth groups of potentials are more often and are
more intensively used for the description of properties for
deuteron and character of his interaction with easy nucleus.

On Fig. 1 is shown interest of researchers to deuteron and to its properties
according to the quoted literature in this article. Obvious not fading
interest. It is connected first of all to studying those processes and
interactions where the direct participant is deuteron. And knowledge its DWF
is necessary for a substantiation and an explanation of corresponding
models. Thus it is necessary to interpret the received experimental data, in
particular tensor polarization.

\pdfximage width 135mm {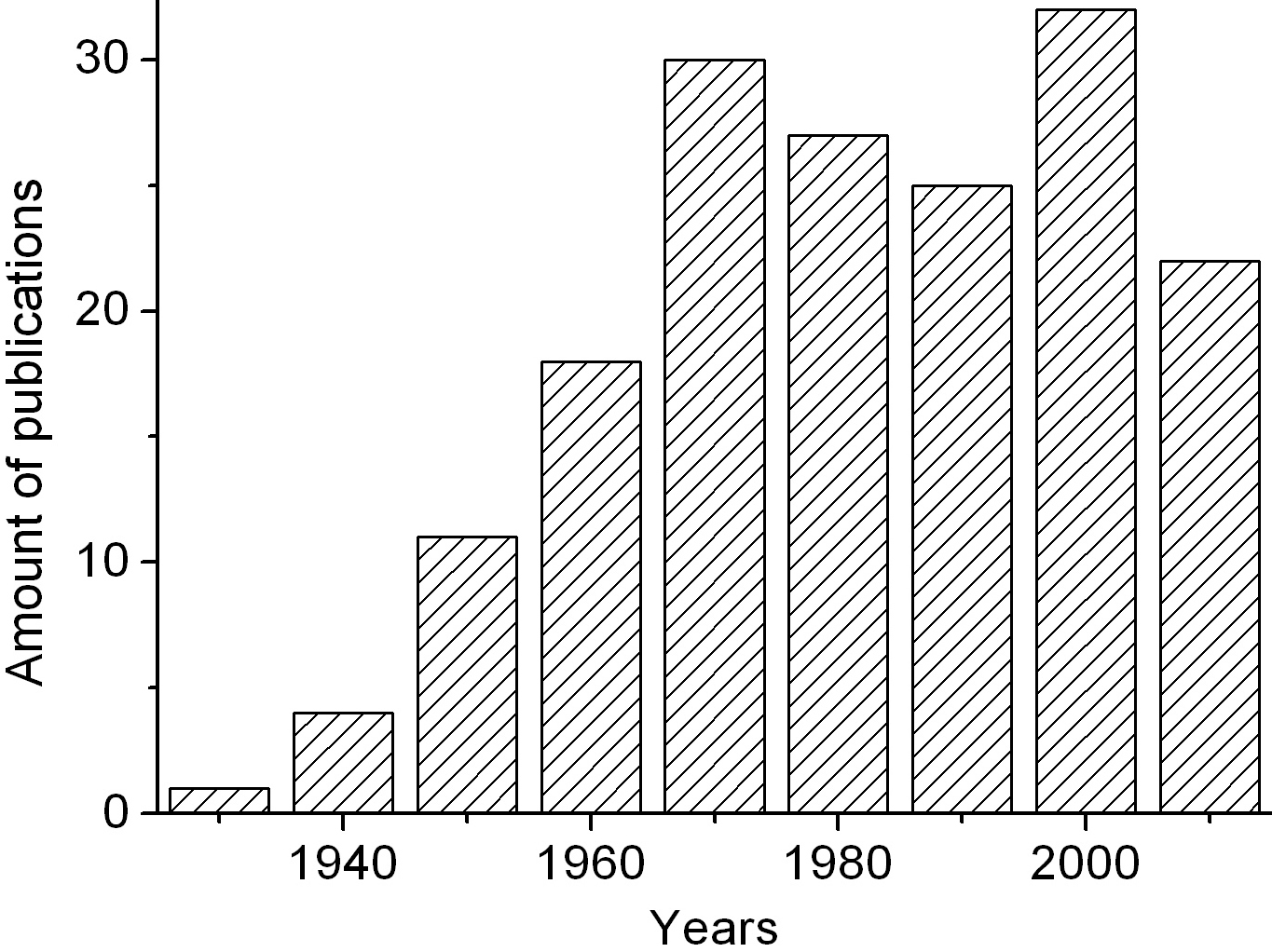}\pdfrefximage\pdflastximage

Fig. 1. Interest of researchers to deuteron

\textbf{3. Deuteron properties}

Based on the known DWFs one can calculate the deuteron properties:

deuteron radius $r_{m}$

\[
r_d = \frac{1}{2}\left\{ {\int\limits_0^\infty {r^2\left[ {u^2(r) + w^2(r)}
\right]dr} } \right\}^{1 / 2};
\]

the quadrupole moment $Q_{d}$

\[
Q_d = \frac{1}{20}\int\limits_0^\infty {r^2w(r)\left[ {\sqrt 8 u(r) - w(r)}
\right]dr} ;
\]

the magnetic moment $\mu _{d}$

\[
\mu _d = \mu _s - \frac{3}{2}(\mu _s - \frac{1}{2})P_D ;
\]

the D- state probability $P_{D}$

\[
P_D = \int\limits_0^\infty {w^2(r)dr} ;
\]

the ``D/S- state ratio'' $\eta $

\[
\eta = A_D / A_S ;
\]

the triplet effective range $\rho $.

In a formula for \textit{$\mu $}$_{d}$ size $\mu _s = \mu _n + \mu _p $ is the sum of the
magnetic moments of a neutron and proton. Value of the calculated magnetic
moment of a deuteron is given in nuclear magnetons \textit{$\mu $}$_{N}$.

Values of these static properties for deuteron that were designed for
different potential models or wave functions of a various origin are
resulted in Table 1. Knots for radial DWFs $u(r)$ and $w(r)$ are designated as $r_{u}$ and
$r_{w}$.

Table 1. Deuteron properties

\begin{tabular}
{|p{0.7cm}|p{3.7cm}|p{0.7cm}|p{0.7cm}|p{1.4cm}|p{1.1cm}
|p{1.3cm}|p{1cm}|p{1.1cm}|p{1.1cm}|p{0.8cm}|}\hline
Years&Potential or DWF& $r_{u}$ (fm)& $r_{w}$ (fm)& $E_{d}$ (MeV)&
$r_{m}$ (fm)& $Q_{d}$ (fm$^{2})$& $P_{D}$ ({\%})& $\eta $& A$_{S}$
(fm$^{-1/2})$&
Ref. \\
\hline 1940& Neutral theory (zero cut-off)& -& -& & & 0.270& 6.8&
& &
\cite{Bethe1940}\\
\hline 1940& Neutral theory (straight cut-off)& -& -& & & 0.261&
6.63& & &
\cite{Bethe1940} \\
\hline 1941& Results of Rarita-Schwinger& -& -& 2.17& & & 3.9& & &
\cite{Rarita1941} \\
\hline 1954& Results of Brueckner-Watson ($V_{T}$=-500 MeV)& & & &
1.97& 0.325& 7.60& & &
\cite{Matsumoto1954} \\
\hline 1954& ... ($V_{T}$=300 MeV)& & & & 1.86& 0.277& 5.10& & &
\cite{Matsumoto1954}\\
\hline 1955& Trial functions& & & 2.227& & 0.28& 7.11& & &
\cite{Cap1955}\\
\hline 1955& Gartenhaus DWF& -& -& & & 0.29-0.308& 6.8-7.0& & &
\cite{Gartenhaus1955} \\
\hline 1956& Pion-theoretical wave function& 0.4& 0.4& & & 0.28&
5-8& 0.0245& &
\cite{Iwadare1956}\\
\hline 1956& Variational wave function& -& -& & & & 17& & &
\cite{noyes1956} \\
\hline 1958& Hulthen type DWF& -& & & & & & & &
\cite{Moravcsik1958} \\
\hline 1959& GT-Potential& 0.4& 0.4& 2.288& & 0.263& 6.3& & &
\cite{laurikainen1959} \\
\hline 1960& Hamada& -& -& & 1.7& 0.273& 6.7& 0.0258& &
\cite{Hamada1960} \\
\hline 1960& Pion-theoretical DWF& -& 0.15& & & 0.26& 7& & &
\cite{Matsumoto1960} \\
\hline 1961& Hamada& -& -& & & & 9.9& 0.029& &
\cite{Hamada1961} \\
\hline 1962& Hamada-Johnston& -& -& 2.226& & 0.285& 6.97& 0.02656&
&
\cite{Hamada1962} \\
\hline 1963& Martin's method& & & & & 0.137& 4& & &
\cite{Bialkowski1963} \\
\hline 1964& Hulthen wave function& 0.15& -& & & & & & &
 \cite{Kottler1964}\\
\hline 1964& separable potential& -& -& 2.225& & & 3.2& & &
 \cite{Tabakin1964}\\
\hline 1966& Hamada-Johnston (analytic)& -& -& & & 0.282& 7&
0.0269& &
\cite{McGee1966}  \\
\hline 1966& Hamada-Johnston (Hulthen)& -& -& & & & & & &
 \cite{McGee1966} \\
\hline 1966& Hamada-Johnston-Partovi& 0.5& 0.5& & & & & & &
 \cite{McGee1966} \\
\hline 1966& Soft core& -& & 2.227& & & & & &
\cite{Eikemeier1966} \\
\hline 1968& Relative harmonic oscillator basis& -& -& 2.1& &
0.325& 3.6& & &
 \cite{elliott1968}\\
\hline 1968& Effective nucleon-nucleon potential (A, B, F
variants)& & & 1.99; \par 2.20; \par 2.13& & 0.272; \par 0.266;
\par 0.227& 1.94; \par 1.97; \par 2.59& & &
\cite{nestor1968} \\
\hline 1968& Soft core Reid68& 0.01& 0.01& 2.2246& & 0.27964&
6.4696& 0.02622& 0.87758&
 \cite{Reid1968}\\
 \hline  1968& Hard core Reid68& 0.38& 0.38& 2.2246& & 0.277&
6.497& 0.0259& 0.88034&
 \cite{Reid1968}\\
\hline 1969& Non-static OBEP (set 1)& -& -& 2.2& & 0.26& 6.3& & &
\cite{erkelenz1969} \\
\hline
\end{tabular}

\begin{tabular}
{|p{0.7cm}|p{3.7cm}|p{0.8cm}|p{0.8cm}|p{1.4cm}|p{1.1cm}
|p{1.3cm}|p{1cm}|p{1.1cm}|p{1.1cm}|p{0.8cm}|}\hline
 1969& ... (set 2)& & & 2.3& & 0.25& 5.4& & &
\cite{erkelenz1969}\\
\hline 1970& Modified HJ v1& 0.4& 0.4& 2.226& & 0.2845& 6.953&
0.02642& &
\cite{Humberston1970} \\
\hline 1970& Modified HJ v3& 0.4& 0.4& 2.2256& & 0.2867& 6.964&
0.02674& &
\cite{Humberston1970} \\
\hline 1970& Modified HJ v9& 0.4& 0.4& 2.2680& & 0.2869& 7.050&
0.02768& &
 \cite{Humberston1970}\\
\hline 1971& Velocity dependent potentials from the various
models: distributed mass scalar& -& -& 2.224& & 0.275& 4.6& & &
 \cite{stagat1971}\\
\hline 1971& L$^{2}$ force& -& -& 2.224& & 0.262& 4.0& & &
\cite{stagat1971} \\
\hline 1971& Contact term& -& -& 2.224& & 0.258& 4.9& & &
\cite{stagat1971} \\
\hline 1971& Phenomenological charge dependent& -& -& 2.224& &
0.240& 4.1& & &
 \cite{stagat1971}\\
\hline 1972& OBEP& & & 2.2& & 0.26& 6.3& & &
\cite{Holinde1972} \\
\hline 1973& Local nucleon-nucleon potential A& & & 2.224& &
0.262& 4.43& & &
 \cite{tourreil1973}\\
\hline 1973& ... B& & & 2.224& &
0.262& 5.25& & &
 \cite{tourreil1973}\\
\hline 1973& ... C& & & 2.224& &
0.279& 5.45& & &
 \cite{tourreil1973}\\
\hline 1973& UT101& 0.6; 0.8& 0.6; 0.8& & & 0.279& & & &
\cite{Vary1973} \\
\hline 1973& UT102& 0.7& 0.7& & & 0.279& & & &
 \cite{Vary1973}\\
\hline 1973& UT103& 0.6; 0.9& 0.6; 0.9& & & 0.279& & & &
 \cite{Vary1973}\\
\hline 1974& Boundary condition model& & & 2.2262& & 0.2774& 5.20&
0.02617& 0.8858&
 \cite{arenhovel1974}\\
\hline 1974& Reid hard core& & & 2.2247& & 0.2769& 6.49& 0.02584&
0.8774&
\cite{arenhovel1974} \\
\hline 1974& Yale & & & 2.1939& & 0.2757& 6.95& 0.02505& 0.8804&
\cite{arenhovel1974} \\
\hline 1974& Hamada-Johnston& & & 2.2710& & 0.2837& 7.02& 0.02686&
0.8921&
\cite{arenhovel1974} \\
\hline 1974& Bryan-Scott potential& & & 2.1841& & 0.2589& 5.44&
0.02375& 0.8687&
 \cite{fabian1974}\\
\hline 1974& Ueda-Green I potential& & & 1.9556& & 0.2811& 5.47&
0.02291& 0.8455&
\cite{fabian1974} \\
\hline 1974& Ueda-Green I potential& & & 2.2052& & 0.2797& 6.01&
0.02567& 0.8881&
 \cite{fabian1974}\\
\hline 1974& Ueda-Green III potential& & & 2.5315& & 0.2605& 4.93&
0.02817& 0.9349&
 \cite{fabian1974}\\
\hline 1974& Separable potential& & & 2.223& & 0.288& 7& 0.0437& &
\cite{pieper1974} \\
\hline 1975& Approximation for Yale potential& & & 2.1888& &
0.276& 6.95& & &
 \cite{afnan1975}\\
\hline 1975& RSC& -& -& & & 0.280& 6.47& & &
\cite{Coester1975} \\
\hline 1975& RHC& 0.5& 0.5& & & 0.277& 6.50& & &
 \cite{Coester1975}\\
\hline 1975& HJ potential& 0.5& 0.5& & & 0.284& 6.95& & &
 \cite{Coester1975}\\
\hline 1975& RHC+Baker transf. of u(r)& -& & & & 0.276& 6.50& & &
\cite{Coester1975}\\
\hline1975& RSC+u-w twist& -& 1.2& & & 0.268& 4.35& & &
 \cite{Coester1975}\\
\hline
\end{tabular}

\begin{tabular}
{|p{0.7cm}|p{3.7cm}|p{0.7cm}|p{0.7cm}|p{1.4cm}|p{1.1cm}
|p{1.3cm}|p{1cm}|p{1.1cm}|p{1.1cm}|p{0.8cm}|} \hline 1975&
RSC+UT101& 0.8& 0.8& & & 0.279& 6.47& & &
\cite{Coester1975} \\
\hline 1975& OBEP HM& & & 2.224& 1.86& 0.284& 5.75& & &
 \cite{Holinde1975}\\
\hline 1975& OBEP SCH& & & 2.910& 1.79& 0.249& 4.85& & &
\cite{Holinde1975} \\
\hline 1975& OBEP GTG& & & 2.985& 1.76& 0.252& 4.88& & &
 \cite{Holinde1975}\\
\hline 1975& OBEP UNG& & & 2.511& 1.81& 0.266& 4.40& & &
 \cite{Holinde1975}\\
\hline 1975& Refitted OBEP SCH'& & & 2.224& 1.85& 0.284& 5.82& & &
 \cite{Holinde1975}\\
\hline 1975& Refitted OBEP GTG'& & & 2.223& 1.85& 0.296& 6.10& & &
 \cite{Holinde1975}\\
\hline 1975& Refitted OBEP GTG''& & & 2.227& 1.82& 0.285& 5.67& &
&
\cite{Holinde1975} \\
\hline 1975& Meson exchange model F$_{0}$F$_{1}$'& -& -& 2.227& &
& 6.17& & &
\cite{jackson1975} \\
\hline 1975& One-boson-exchange potential& 0.48& 0.48& 2.224644& &
& 5.92& 0.0251& &
 \cite{nagels1975}\\
\hline 1975& OBEH(R)& 0.4& -& 2.231& & 0.2747& 6.23& & &
 \cite{Obinata1975}\\
\hline 1975& OBEH(NR)& & & 2.232& & 0.2721& 5.57& & &
 \cite{Obinata1975}\\
\hline 1975& OBEG(R)& -& -& 2.227& & 0.2740& 6.14& & &
 \cite{Obinata1975}\\
\hline 1975& OBEG(NR)& & & 2.205& & 0.2720& 5.58& & &
\cite{Obinata1975} \\
\hline 1975& OBEV(R)& -& -& 2.205& & 0.2745& 5.63& & &
\cite{Obinata1975} \\
\hline 1975& OBEV(NR)& & & 2.244& & 0.2698& 5.23& & &
\cite{Obinata1975} \\
\hline 1975& Super-soft-core potential& & & 2.2245& & 0.282& 5.92&
& &
\cite{tourreil1975} \\
\hline 1976& OBEP Holinde-Machleidt model& -& -& 2.224& 1.86&
0.284& 5.75& & &
 \cite{Holinde19761}\\
\hline 1976& OBEP Holinde-Machleidt model& & & 2.2246& 1.79&
0.2864& 4.32& & &
\cite{Holinde19762} \\
\hline 1976& Exact, Kim-Vasavada's, Brysk- Michalik's DWF& -& & &
& & & & &
 \cite{Weiss1976} \\
\hline 1977& Analytic wave function & -& -& & & 0.288& 4& & &
\cite{Adler1977} \\
\hline 1977& RSC potential with pion Compton wavelength& 0.3& 0.3&
& & 0.2732- \par 0.2798& 4.5-6.5& & &
 \cite{mcgurk1977}\\
\hline 1978& Analytic wave function & 0.25-0.5& 0.3-0.5& & & & & &
&
 \cite{Allen1978}\\
\hline 1978& KLS& -& -& 2.16& & 0.093& 0.32& & &
\cite{Mathelitsch1978} \\
\hline 1978& Graz I& -& -& 2.225& & 0.288& 2.63& & &
\cite{Mathelitsch1978} \\
\hline 1978& Mongan II& -& 1.2& 2.223& & 0.275& 1.12& & &
\cite{Mathelitsch1978} \\
\hline 1978& Low-energy nucleon-nucleon potential from Regge-pole
theory& -& -& & & 0.2775& 5.39& 0.0255& 0.8015&
 \cite{nagels1978}\\
\hline 1979& Interactions in the core region& 0.5& 0.5& & & 0.279&
5.45& & &
\cite{allen1979} \\
\hline 1979& Super soft-core potential& -& -& & & 0.279& 5.45& & &
 \cite{allen1979}\\
\hline 1979& OBE ($\lambda $=0)& 0.2& -& & & & 4.74& & &
 \cite{Buck1979}\\
\hline 1979& OBE ($\lambda $=0.4)& 0.2& -& & & & 4.78& & &
\cite{Buck1979} \\
\hline 1979& OBE ($\lambda $=1.0)& 0.25& 0.5& & & & 3.60& & &
 \cite{Buck1979}\\
\hline 1979& OBEP model& & & 2.22464& & 0.284& 6.36& 0.0261&
0.797&
\cite{nagels1979} \\
\hline
\end{tabular}

\begin{tabular}
{|p{0.7cm}|p{3.7cm}|p{0.7cm}|p{0.7cm}|p{1.4cm}|p{1.1cm}
|p{1.3cm}|p{1cm}|p{1.1cm}|p{1.1cm}|p{0.8cm}|} \hline 1980& Paris potential&
-& -& 2.2249& & 0.279& 5.77& 0.02608& &
\cite{Lacombe1980} \\
 \hline 1980&
Four-component relativistic models& 0.3& 0.2-0.6& & & & & & &
\cite{Arnold1980} \\
\hline 1980& S potential& & & & & 0.286& 6.7& 0.026& &
 \cite{lamot1980}\\
\hline 1980& SF potential& & & & & 0.285& 4.0& 0.027& &
 \cite{lamot1980}\\
\hline 1980& QT interactions& & & & & 0.352& 4.1& 0.038& &
 \cite{lamot1980}\\
\hline 1981& YY7& & & & 1.722& 0.283& 7.0& 0.029& &
 \cite{Koike1981}\\
\hline 1981& YY4& & & & 1.723& 0.283& 4.0& 0.029& &
\cite{Koike1981} \\
\hline 1981& T4D-2 & & & & 1.744& 0.282& 4.0& -0.004& &
\cite{Koike1981} \\
\hline 1981& T4D-1& & & & 1.201& 0.282& 4.0& -0.004& &
\cite{Koike1981} \\
\hline 1981& Urbana potential& & & 2.225& & 0.273& 5.2& 0.025& &
\cite{lagaris1981} \\
\hline 1984& PEST potential& & & 2.2249& & 0.279& 5.77& 0.0261& &
\cite{haidenbauer1984} \\
\hline 1984& FSP& 0.5& 0.5& 2.2246& 1.9549& 0.2727& 6.315&
0.02544& 0.8766&
 \cite{kukulin1984}\\
\hline 1984& Mehdi-Gupta parametrization (shape-1)& & & & &
0.1978- \par 0.2745& 2-6& & &
 \cite{Mehdi1984}\\
\hline 1984& Mehdi-Gupta parametrization (shape-2)& & & & &
0.2252- \par 0.2813& 2-6& & &
\cite{Mehdi1984} \\
\hline 1984& Argonne v14 & -& -& 2.2250& & 0.286& 6.08& 0.0266&
0.845&
 \cite{Wiringa1984}\\
\hline 1984& Argonne v28 & -& -& 2.2250& & 0.286& 6.13& 0.0265&
0.846&
\cite{Wiringa1984} \\
\hline 1985& Realistic superdeep local NN-potential (Moscow)&
0.55& 0.55& 2.2246& 1.9611& 0.2860& 6.78& 0.0269& 0.8814&
 \cite{Krasnopolsky1985}\\
\hline 1986& BEST potential& & & 2.225 & & 0.2855& 4.58& 0.0267&
0.8950&
 \cite{haidenbauer1986}\\
\hline 1986& Quark compound bag model (b=1.2 fm)& & & & & & 5.33&
0.02609& 0.8945&
 \cite{Kalashnikova1986}\\
\hline 1986& ... (b=1.4 fm)& & & & & & 4.66& 0.02609& 0.8757&
 \cite{Kalashnikova1986}\\
\hline 1986& ... (b=1.6 fm)& & & & & & 4.26& 0.02609& 0.8884&
\cite{Kalashnikova1986} \\
\hline 1986& Positive short range tensor model potential& -& 0.8&
2.22464& 1.9726& 0.2860& & 0.02639& 0.8847&
\cite{kermode1986} \\
\hline 1987& NN potentials with six-quark core radius b=1fm& & &
2.22462& 1.96& 0.276& 5.7& 0.0258 & &
 \cite{beyer1987}\\
\hline 1987& ... b=1.2fm& & & 2.22462& 1.99& 0.286& 5.3& 0.0263& &
 \cite{beyer1987}\\
\hline 1987& Certov- Mathelitsch- Moravcsik DWF& up 0.1& up 0.1& &
1.959- \par 1.975& 0.280& 4;6;8& 0.0261& 0.88688&
\cite{Certov1987} \\
\hline 1987& Microscopic meson-quark cluster model (set A)& -& -&
& & 0.266& 5.23& & &
 \cite{ito1987}\\
\hline 1987& ... (set B)& -& -& & & 0.268& 5.33& & &
 \cite{ito1987}\\
\hline 1987& OBEP full model& -& 0.3& 2.2246& 2.0016& 0.2807&
4.249& 0.02668& 0.9046&
 \cite{Machleidt1987}\\
\hline 1987& OBEPQ& -& 0.04; \par 0.5& 2.2246& 1.9684& 0.274&
4.38& 0.0262& 0.8862&
\cite{Machleidt1987} \\
\hline 1987& OPE& 0.25& 0.25& & & & 6& 0.0262& &
\cite{Righi1987} \\
\hline
\end{tabular}

\begin{tabular}
{|p{0.7cm}|p{3.7cm}|p{0.7cm}|p{0.7cm}|p{1.4cm}|p{1.1cm}
|p{1.3cm}|p{1cm}|p{1.1cm}|p{1.1cm}|p{0.8cm}|}  \hline
 1988&
Nonlocal potential ($\lambda $=5fm$^{ - 3})$& -& -& 2.22448&
1.96880& 0.23953& 4.9989& 0.02198& 0.8861&
 \cite{mustafa1988}\\
\hline 1988& Nonlocal potential ($\lambda $=375fm$^{ - 3})$& 0.5&
0.8& 2.22466& 1.98547& 0.30270& 8.8181& 0.02570& 0.8856&
 \cite{mustafa1988}\\
\hline 1988& Phenomenological realistic DWF& & & & 1.953& 0.286& &
0.0268& 0.8800&
 \cite{oteo1988}\\
\hline 1989& OBEPA& -& 0.05; \par 0.4& 2.22452& 1.9693& 0.274&
4.38& 0.0263& 0.8867&
\cite{Machleidt1989}\\
\hline 1989& OBEPB& -& 0.02& 2.22461& 1.9688& 0.278& 4.99& 0.0264&
0.8860&
\cite{Machleidt1989} \\
\hline 1989& OBEPC& -& 0.01& 2.22459& 1.9674& 0.281& 5.61& 0.0266&
0.8850&
\cite{Machleidt1989} \\
\hline 1989& Quark compound bag model QCB82& 0.4& 0.4& 2.224574& &
0.2777& 5.34& 0.02593& 0.8891&
\cite{dijk1989} \\
\hline 1989& ... QCB86& 0.6& 0.6& 2.224574& & 0.2786& 5.47&
0.02597& 0.8894&
 \cite{dijk1989}\\
\hline 1990& Quark cluster model (set A and B)& -& -& & & & 5.4;
4.9& & &
\cite{buchmann1990} \\
\hline 1990& Quark compound bag model (b=1.2 fm)& & & 2.2249&
1.9725& 0.279& 5.30& 0.0261& 0.8874&
\cite{grach1990} \\
\hline 1990& Quark compound bag model (b=1.35 fm)& & & 2.2249&
1.9751& 0.278& 4.66& 0.0261& 0.8889&
 \cite{grach1990}\\
\hline 1991& Padua potential& -& -& 2.2249& 1.9725& 0.279& 5.3&
0.0261& 0.8874&
 \cite{Minelli1991}\\
\hline 1992& Full folded-diagram potential& & & 2.2244& & 0.2796&
5.22& 0.0264& 0.8886&
\cite{haidenbauer1992}\\
\hline 1992& Moscow \textit{NN } model& 0.65& -& 2.2245& 1.9592&
0.2859& 6.75& 0.0269& &
\cite{kukulin1992} \\
\hline 1993& Nonlocal potential& 0.5& 0.5& 2.2242& 1.953& 0.2862&
6.544& 0.0287& 0.8898&
\cite{mustafa1993} \\
\hline 1993& Coupled-coupled folded-diagram potential& & & 2.2245&
& 0.2852& 5.58& 0.0267& 0.8927&
 \cite{haidenbauer1993}\\
\hline 1994& OPE (R=0.8906313)& -& -& & 1.9366& 0.2751& 5.862&
0.02653& 0.86952&
 \cite{Sprung1994}\\
\hline 1994& Inversion potential& -& -& 2.224579& 1.9702& 0.2816&
5.91& 0.0264& 0.8860&
\cite{Kohlhoff1994} \\
\hline 1994& Nijm-3& -& -& 2.224576& 1.9672& 0.2705& 5.53& 0.0252&
0.8848&
 \cite{Kohlhoff1994}\\
\hline 1994& Quantum inversion by Newton-Fulton
(\textit{original})& -& 1.3& 2.232139& 1.85& 0.275& 2.09&
0.018081& 0.8269&
 \cite{Kohlhoff1994}\\
\hline 1994& Newton-Fulton (\textit{wrong})& -& 1.8& 2.232139&
1.935& 0.0925& 1.00& 0.018071& 0.8753&
\cite{Kohlhoff1994} \\
\hline 1994& Newton-Fulton (\textit{correct})& -& -& 2.232139&
1.947& 0.2310& 6.77& 0.018081& 0.8753&
 \cite{Kohlhoff1994}\\
\hline 1994& Quark cluster model& -& -& 2.2246& 1.9657& & 4.91&
0.0261& 0.8765&
 \cite{valcarce1994}\\
\hline 1994& Nijm I& & & 2.224575& & 0.2719& 5.664& 0.0253&
0.8841&
 \cite{stoks1994}\\
\hline 1994& Nijm II& & & 2.224575& & 0.2707& 5.635& 0.0252&
0.8845&
 \cite{stoks1994}\\
\hline 1994& Reid 93& & & 2.224575& & 0.2703& 5.699& 0.0251&
0.8853&
\cite{stoks1994} \\
\hline 1994& Nijm 93& & & 2.224575& & 0.2706& 5.755& 0.0252&
0.8842&
 \cite{stoks1994}\\
\hline 1995& Complex Kohn variational& -& -& 2.2298& & & &
0.02634& &
\cite{Araujo1995} \\
\hline 1995& OBEPR, OBEPR(A), OBEPR(B)& -& 0.2& & & & & & &
\cite{Levchuk1995} \\
\hline 1995& Argonne v18& -& -& 2.22457& 1.967& 0.270& 5.76&
0.0250& 0.8850&
\cite{Wiringa1995}\\
\hline 1995& NijmI, NijmII, Reid93& -& -& & & & & & &
\cite{Swart1995} \\
\hline
\end{tabular}

\begin{tabular}
{|p{0.7cm}|p{3.7cm}|p{0.7cm}|p{0.7cm}|p{1.4cm}|p{1.1cm}
|p{1.3cm}|p{1cm}|p{1.1cm}|p{1.1cm}|p{0.8cm}|} \hline 1996& SDA& &
& 2.2246& 1.965& 0.275& 3.5948& 0.02715& 0.885&
 \cite{Doleschall1996}\\
\hline 1996& SDB& & & 2.2246& 1.9649& 0.2750& 3.6233& 0.02706&
0.8850&
 \cite{Doleschall1996}\\
\hline 1996& SDC& & & 2.2246& 1.9646& 0.2749& 3.4202& 0.02723&
0.8849&
 \cite{Doleschall1996}\\
\hline 1996& SDD& & & 2.2246& 1.9657& 0.2750& 4.315l& 0.02647&
0.8849&
 \cite{Doleschall1996}\\
\hline 1996& Reid, Paris,Urbana, Argonne v18& -& -& & & & & & &
\cite{forest1996} \\
\hline 1996& Resonating-group method \par (RGM-F)& & & 2.274&
1.933& 0.2752& 5.391& 0.0264& &
 \cite{Fujiwara1996}\\
\hline 1996& FSS& -& -& 2.244& 1.966& 0.2845& 5.879& 0.0272& &
 \cite{Fujiwara1996}\\
\hline 1996& RGM-H& & & 2.224& 1.986& 0.2750& 4.998& 0.0251& &
 \cite{Fujiwara1996}\\
\hline 1996& Effective chiral Lagrangian model fitted values
($\Lambda $=2.5fm$^{ - 1})$& & & 2.15& & 0.246& 2.98& 0.0229& &
 \cite{Ordonez1996}\\
\hline 1996& ... ($\Lambda $=3.9fm$^{ - 1})$& -& 1& 2.24& & 0.249& 2.86& 0.0244& &
 \cite{Ordonez1996}\\
\hline 1996& ... ($\Lambda $=5fm$^{ - 1})$& & & 2.18& & 0.237& 2.4& 0.023& &
\cite{Ordonez1996} \\
\hline 1998& One solitary boson exchange potential (OSBEP)& & &
2.22459& 1.9554& 0.2728& 6.0& 0.0256& 0.8788&
\cite{Jade1998} \\
\hline 1998& Moscow A& 0.5& 0.5& 2.2244& 1.96& & 6.59& 0.0267& &
 \cite{Kukulin1998}\\
\hline 1998& Moscow B& 0.5& 0.5& 2.2246& 1.95& & 5.75& 0.0258& &
\cite{Kukulin1998} \\
\hline 1998& Moscow C& 0.5& 0.5& 2.2246& 1.94& & 6.14& 0.0262& &
\cite{Kukulin1998} \\
\hline 1999& OPE& 0.8& 0.8& 2.224589& 1.965& 0.2859& 5.86& 0.0271&
0.8836&
 \cite{Gridnev1999}\\
\hline 2000& NLO& -& -& 2.1650& 1.975& 0.266& 3.62& 0.0248& 0.866&
 \cite{epelbaum2000}\\
\hline 2000& NNLO& 1.1& -& 2.2238& 1.967& 0.262& 6.11& 0.0245&
0.884&
\cite{epelbaum2000} \\
\hline 2000& NNLO-$\Delta $& & & 2.1849& 1.970& 0.268& 5.00&
0.0247& 0.873&
\cite{epelbaum2000} \\
\hline 2000& Local NN Potential LP1& 0.5-0.52& -& 2.2246& 1.965&
0.271& 5.62& 0.0253& 0.884&
 \cite{Dubovichenko20001}\\
\hline 2000& ... LP2& 0.5-0.52& -& 2.2246& 1.966& 0.274& 5.75&
0.0256& 0.884&
 \cite{Dubovichenko20001}\\
\hline 2000& ... LP3& 0.5-0.52& -& 2.2246& 1.967& 0.279& 6.00&
0.0261& 0.884&
\cite{Dubovichenko20001} \\
\hline 2000& ... LP4& 0.5-0.52& -& 2.2246& 1.968& 0.285& 6.23&
0.0266& 0.884&
 \cite{Dubovichenko20001}\\
\hline 2000& ... LP5& 0.5-0.52& -& 2.2246& 1.968& 0.290& 6.56&
0.0273& 0.884&
\cite{Dubovichenko20001} \\
\hline 2001& Argonne V18& -& -& & & & & & &
 \cite{Garcon2001}\\
\hline 2001& Bonn C& -& -& & 1.968& 0.2814& 5.60& 0.0266& &
\cite{Fujiwara2001} \\
\hline 2001& FSS2 (Isospin basis)& -& -& 2.2250& 1.9598& 0.2696&
5.490& 0.02527& &
 \cite{Fujiwara2001}\\
\hline 2001& FSS2 (Particle basis, Coulomb off)& & & 2.2261&
1.9599& 0.2696& 5.490& 0.02527& &
 \cite{Fujiwara2001}\\
\hline 2001& FSS2 (Particle basis, Coulomb on)& & & 2.2309&
1.9582& 0.2694& 5.494& 0.02531& &
\cite{Fujiwara2001} \\
\hline 2001& CD-Bonn& -& 0.1& 2.224575& 1.966& 0.270& 4.85&
0.0256& 0.8846&
 \cite{Machleidt2001}\\
\hline
\end{tabular}

\begin{tabular}
{|p{0.7cm}|p{3.7cm}|p{0.7cm}|p{0.7cm}|p{1.4cm}|p{1.1cm}
|p{1.3cm}|p{1cm}|p{1.1cm}|p{1.1cm}|p{0.8cm}|}
 \hline 2001&
Separable potentials with the Laguerre form factors& 0.2& -& &
0.2819& & 5.729& 0.0252& 0.8845&
 \cite{Zaitsev2001}\\
\hline 2001& Idaho-A& -& -& 2.224575& 1.9756& 0.281& 4.17& 0.0256&
0.8846&
\cite{Entem2001} \\
\hline 2001& Idaho-B& -& -& 2.224575& 1.9758& 0.284& 4.94& 0.0255&
0.8846&
 \cite{Entem2001}\\
\hline 2003& Nij1 transformed& 0.6& 0.4& & & & & & &
\cite{Amghar2003} \\
\hline 2003& Nij2 transformed& -& -& & & & & & &
 \cite{Amghar2003}\\
\hline 2003& DBS model NN& 0.6& 0.5& 2.22454& 2.004& 0.286& 5.42&
0.0259& 0.9031&
\cite{Kaskulov2003} \\
\hline 2003& DBS model NN + 6q& & & 2.22454& 1.972& 0.275& 5.22&
0.0264& 0.8864&
\cite{Kaskulov2003} \\
\hline 2003& Idaho N3LO (500)& & & 2.224575& 1.978& 0.285& 4.51&
0.0256& 0.8843&
\cite{Entem2003} \\
\hline 2004& Exponential potential& 0.5& -& 2.2246& 1.960& 0.283&
6.22& 0.0265& 0.881&
\cite{Dubovichenko2004} \\
\hline 2004& Modified Moscow& & & 2.22453& 1.956& 0.286& 6.776&
0.0269& 0.879&
\cite{Dubovichenko2004} \\
\hline 2004& ISTP v.0& -& -& 2.224575& 1.9877& & 4.271& 0.0252&
0.8845&
\cite{Shirokov2004} \\
\hline 2004& ISTP v.1& -& -& 2.224575& 1.9997& & 5.620& 0.0252&
0.8845&
\cite{Shirokov2004} \\
\hline 2004& ISTP v.2& -& -& 2.224575& 1.9680& & 5.696& 0.0252&
0.8629&
\cite{Shirokov2004} \\
\hline 2005& OPE-$\eta $ (LO)& & & & 1.9423& 0.1321& 6& 0& 0.8752&
\cite{Valderrama2005} \\
\hline 2005& OPE-pert (NLO)& & & & 1.6429& 0.4555& 0& 0.051&
0.7373&
 \cite{Valderrama2005}\\
\hline 2005& OPE-exact& up 0.5& up 0.5& & 1.9351& 0.2762& 7.88&
0.02633& 0.8681&
\cite{Valderrama2005} \\
\hline 2005& Nonrelativistic DWF& -& -& 2.2245& 2.108& 0.2859& & &
&
\cite{Berezhnoy2005} \\
\hline 2005& NLO& -& -& 2.171- \par 2.186& 1.973- \par 1.974&
0.273- \par 0.275& 3.46- \par 4.29& 0.0256- \par 0.0257& 0.868-
\par 0.873&
 \cite{Epelbaum2005}\\
\hline 2005& NNLO& -& -& 2.189-2.202& 1.970-1.972&
0.271- \par 0.275& 3.53-4.93& 0.0255-0.0256& 0.874-
\par 0.879&
 \cite{Epelbaum2005}\\
\hline 2005& N3LO& 0.5& -& 2.216-2.223& 1.973-1.985&
0.264- \par 0.268& 2.73-3.63& 0.0254-0.0255& 0.882-0.883&
 \cite{Epelbaum2005}\\
\hline 2006& Moscow& 0.5& -& 2.2246& 1.9639& 0.2674& & 0.02714&
0.8892&
 \cite{Knyr2006}\\
\hline 2007& Moscow& 0.5& -& & & & & & &
\cite{Khokhlov2007} \\
\hline 2007& Renormalized OPE and TPE chiral potentials& up 0.5&
up 0.5& & & & & 0.02633; \par 0.02564& &
\cite{Arriola2007} \\
\hline 2007& MT wave function& -& -& 2.224996& 1.972& 0.2731& 6.2&
0.0253& &
\cite{Krutov2007}  \\
\hline 2007& JISP16& -& -& 2.224576& 1.9647& 0.2915& 4.136&
0.0252& 0.8629&
 \cite{Mazur2007}\\
\hline 2008& LO $\chi $ET, NNLO $\chi $ET& up 0.5& up 0.6& & 1.90-
\par 2.06& 0.276- \par 0.359& 6.98- \par 10.08& 0.0251- \par
0.0302& 0.845- \par 0.925&
 \cite{Valderrama2008}\\
\hline 2008& OPE& 0.45& 0.5& 2.224575& 1.9351& 0.2762& 7.88&
0.02634& 0.8681&
 \cite{Higa2008}\\
\hline 2008& HB-TPE set IV& & & 2.224575& 1.967& 0.276& 8& Input&
0.884&
 \cite{Higa2008}\\
\hline 2008& RB-TPE set IV& 0.1-0.6& 0.55& 2.224575& 1.8526&
0.3087& 22.99& 0.03198& 0.8226&
\cite{Higa2008} \\
\hline 2008& RB-TPE set $\eta $& -& 0.5-0.8& 2.224575& 1.96776&
0.2749& 5.59& 0.02566& 0.88426&
\cite{Higa2008} \\
\hline 2009& NNLO& 0.5& 0.5& & & & & & &
\cite{Yang2009} \\
\hline 2009& LO& 0.2; \par 0.5& 0.2; \par 0.5& & 1$.$9351&
0$.$2762& 7$.$31& 0$.$02633& 0$.$8681&
\cite{Valderrama2009} \\
\hline
\end{tabular}

\begin{tabular}
{|p{0.7cm}|p{3.7cm}|p{0.7cm}|p{0.7cm}|p{1.4cm}|p{1.1cm}
|p{1.3cm}|p{1cm}|p{1.1cm}|p{1.1cm}|p{0.8cm}|} \hline 2009&
NLO-$\Delta $& 0.1-0.6& 0.1-0.8& & 1$.$963& 0$.$274& 5$.$9& &
0$.$884&
 \cite{Valderrama2009}\\
\hline 2009& N2LO-$\Delta $& 0.1-0.7& 0.1-0.6& & 1$.$980& 0$.$279&
5$.$9& & 0$.$892&
\cite{Valderrama2009} \\
\hline 2010& Oxford potential& & & 2.2246& 1.9767& 0.2871& 5.604&
0.0262& 0.8918&
 \cite{downum2010}\\
\hline 2011& GWU PWA& & & 2.224575& 1.9557& 0.2852& & 0.0256&
0.8764&
 \cite{Babenko2011}\\
\hline 2011& Nijm PWA93& & & 2.224575& 1.9673& 0.2884& & 0.0256&
0.8845&
\cite{Babenko2011} \\
\hline 2011& Yakawa Potential& -& & 2.228& & & & & &
 \cite{Shojaei2011} \\
\hline 2012& Hulthen wave function& -& -& & & & & & &
\cite{Lamia2012} \\
\hline 2013& $\delta$ shell potential& & & & 1.9645& 0.2679& 5.62&
0.02493& 0.8829&
\cite{Perez2013} \\
\hline 2014& DWF in continuum basis& -& -& 2.210& & & 6.3& & &
 \cite{Betan2014}\\
\hline 2014& Coarse-grained NN potential with chiral two-pion
exchange& & & & 1.9689& 0.2658& 5.30& 0.02473& 0.8854&
\cite{Perez20141}\\
\hline 2014& Statistical error analysis for potentials& & & &
1.9744& 0.2645& 5.30& 0.02448& 0.8885&
\cite{Perez20142} \\
\hline 2014& Standard Wood-Saxon potential& -& -& & 1.9532& 0.2769&
6.659& & &
\cite{rezaei2014} \\
\hline 2014& Generalized Wood-Saxon potential& & & & 1.7269& 0.2818&
5.056& & &
\cite{rezaei2014} \\
\hline 2014& Modified Wood-Saxon potential& & & & 1.9532& 0.2836& 4.86&
& &
\cite{rezaei2014} \\
\hline 2015& Idaho N3LO (500)& -& -& 2.2246& 1.975& 0.275& 4.51&
0.0256& 0.8843&
\cite{Epelbaum2015}  \\
\hline 2015& Juelich N3LO (550/600)& -& -& 2.2196& 1.977& 0.266&
3.28& 0.0254& 0.8820&
 \cite{Epelbaum2015} \\
\hline 2015& Improved N3LO (R=0.8fm)& 0.5& 0.8& 2.2246& 1.970&
0.268& 3.78& 0.0255& 0.8843&
 \cite{Epelbaum2015} \\
\hline 2015& ... (R=0.9mm)& 0.5& -& 2.2246& 1.972& 0.271& 4.19&
0.0255& 0.8845&
\cite{Epelbaum2015}  \\
\hline 2015& ... (R=1.0fm)& -& -& 2.2246& 1.975& 0.275& 4.77&
0.0256& 0.8845&
 \cite{Epelbaum2015} \\
\hline 2015& ... (R=1.1fm)& -& -& 2.2246& 1.979& 0.279& 5.21&
0.0256& 0.8846&
 \cite{Epelbaum2015} \\
\hline 2015& ... (R=1.2fm)& -& -& 2.2246& 1.982& 0.283& 5.58&
0.0256& 0.8846&
  \cite{Epelbaum2015}\\
\hline 2015& FSS2& -& -& 2.2206& 1.961& 0.270& 5.52& 0.0252& &
 \cite{Fukukawa2015}\\
\hline 2015& Nonlocal potentials with chiral TPE including $\Delta
$ resonances. Model a& -& -& 2.224575& 1.948& 0.257& 4.94&
0.0245& 0.8777&
 \cite{piarulli2015}\\
\hline 2015& ... Model b& -& -& 2.224574& 1.975& 0.268& 5.29& 0.0248&
0.8904&
\cite{piarulli2015} \\
\hline 2015& ... Model c& -& -& 2.224575& 1.989& 0.269& 5.55& 0.0246&
0.8964&
 \cite{piarulli2015}\\
\hline
\end{tabular}

Experimental values \cite{Garcon2001, Takigawa2017} of static
properties for deuteron it is specified in Table 2.

Table 2. Experimental properties for deuteron

\begin{tabular}
{|p{131pt}|p{100pt}|p{60pt}|}
\hline
Properties&
Values&
Ref. \\
\hline
Spin&
1&
 \\
\hline
Mean life&
Stable&
 \\
\hline
Mass (u)&
2.01410219(11)&
\cite{Takigawa2017} \\
\hline
Mass (MeV)&
1875.61282(16)&
\cite{Takigawa2017} \\
\hline
Magneticm moment (\textit{$\mu $}$_{N})$&
0.8574382308(72)&
\cite{Takigawa2017} \\
\hline
$E_{d}$ (MeV)&
2.22456612(48)&
\cite{Garcon2001} \\
\hline
$r_{m}$ (fm)&
1.975(3)&
\cite{Garcon2001} \\
\hline
$Q_{d}$ (fm$^{2})$&
0.2859(3)&
\cite{Garcon2001} \\
\hline
$\eta $&
0.0256(4)&
\cite{Garcon2001} \\
\hline
\end{tabular}

According to the General mathematical theorem on the number of
knots of eigenfunctions of boundary value problems
\cite{Courant1953} the function describing the ground state of the
particle becomes zero only at the ends of the interval, and inside
it she knots will have.

In paper \cite{Dubovichenko2004} S.B.Dubovichenko considering the
possibility of the existence of knots VFD. If we consider the
deuteron as a six-quark system, in accordance with a generalized
Levinson theorem \cite{Neudatchin1974, Kukulin1979} triplet S
phase scattering starts with 360$^{o}$ and singlet with 180$^{o}$
up 220$^{o}$. In the D wave, there is a single bound state is
enabled, which, together with the S wave determines the ground
state of the deuteron.

Therefore, the availability of knots due to the numerical calculations or
used potential model.

4. \textbf{Analytical forms of DWF in the years 1939-1969}

When describing DWF in the coordinate representation using terms such as
``analytical shape (form)'', its ``approximation'' or ``parameterization''.
Familiar in the first place, the term ``analytical form'' is used as the
obtained solution of a system of coupled equations. Later in the works is a
expression used to refer to records HFD resulting approximation.

Analytical forms of deuteron wave function are provided with use according
to the designations specified in the quoted literature.

The work written by Flugge \cite{flugge1939}  in 1939 was one of
the first works on research of a deuteron and its quadruple
moment. For calculations such deuteron functions for S- and D-
states were used

\[
\psi _S = \frac{(\alpha b)^{3 / 2}}{\sqrt {8\pi } }\exp \left( { -
\frac{1}{2}\alpha br} \right),
\]

\[
\psi _D = \frac{(\alpha b)^{7 / 2}}{\sqrt {2176\pi } }r^2\exp \left( { -
\frac{1}{2}\alpha br} \right),
\]

\noindent
where $a$=1.3 and $\alpha $=1.34.

H.A. Bethe  \cite{Bethe1940} was one of the first considered the
deuteron is a mixture of $^{3}$S$_{1}$ and $^{3}$D$_{1}$ state.
Then the complete wave function is

\[
\psi = \frac{1}{r}\left[ {\chi (r)F_{10M} + \varphi (r)F_{12M} } \right],
\]

where $F_{JLM}$ are the angular functions; $\chi $ and $\phi $ are the radial
wave functions of the S and D component.

The radial deuteron wave functions satisfy the two coupled differential
equations

\[
\begin{array}{l}
 \frac{d^2\chi }{dr^2} = A\chi - B\sqrt 2 \varphi , \\
 \frac{d^2\varphi }{dr^2} = \left( {A + B + \frac{6}{r^2}} \right)\,\varphi
- B\sqrt 2 \chi , \\
 \end{array}
\]

where

\[
\begin{array}{l}
 A = ae^{ - r} / 3r + \varepsilon ^2, \\
 B = ae^{ - r}\left( {\frac{1}{r^3} + \frac{1}{r^2} + \frac{1}{3r}} \right).
\\
 \end{array}
\]

The potential was be cut off at small distances, therefore we consider the
two alternatives:

1) Zero cut-off ($r < r_0 )$

\[
A = B = 0;
\]

2) Straight cut-off ($r < r_0 )$

\[
\begin{array}{l}
 A = A_0 = ae^{ - r_0 } / 3r_0 + \varepsilon ^2; \\
 B = B_0 = ae^{ - r_0 }\left( {\frac{1}{r_0 ^3} + \frac{1}{r_0 ^2} +
\frac{1}{3r_0 }} \right). \\
 \end{array}
\]

The outside solution is pairs

\[
\begin{array}{l}
 \chi _1 = e^{ - \varepsilon z} + \frac{a}{6\varepsilon }G(r), \\
 \varphi _1 = - \frac{\sqrt 2 a}{8\varepsilon ^4}\left\{ {\left[
{\frac{2\varepsilon ^3 + 3\varepsilon - 3}{r^2} + \frac{(2\varepsilon -
1)\varepsilon ^2}{r}} \right]e^{ - (1 + \varepsilon )r} + \frac{3 -
4\varepsilon ^2}{2\varepsilon }F(r)} \right\}; \\
 \end{array}
\]

\[
\begin{array}{l}
 \chi _2 = - \frac{\sqrt 2 a}{12}\left\{ {\left[ {\frac{1}{r^3} + \frac{1 +
\varepsilon }{r^2} + \frac{2\varepsilon - 1}{r}} \right]e^{ - (1 +
\varepsilon )r} + \frac{3 - 4\varepsilon ^2}{12\varepsilon }G(r)} \right\},
\\
 \varphi _2 = \left( {\frac{1}{r^2} + \frac{\varepsilon }{r} +
\frac{\varepsilon ^2}{3}} \right)e^{ - \varepsilon r} + a\left\{ {H(r)e^{ -
(1 + \varepsilon )r} + \left[ {\frac{3}{32\varepsilon ^5} -
\frac{5}{16\varepsilon ^3} + \frac{1}{12\varepsilon }} \right]F(r)}
\right\}; \\
 \end{array}
\]

where

\[
\begin{array}{l}
 G(r) = - e^{ - \varepsilon r}Ei\left( { - r} \right) + e^{\varepsilon
r}Ei\left( { - r(1 + 2\varepsilon )} \right), \\
 F(r) = - e^{ - \varepsilon r}Ei\left( { - r} \right)\left( {\frac{1}{r^2} +
\frac{\varepsilon }{r} + \frac{\varepsilon ^2}{3}} \right) + e^{\varepsilon
r}Ei\left( { - r(1 + 2\varepsilon )} \right)\left( {\frac{1}{r^2} -
\frac{\varepsilon }{r} + \frac{\varepsilon ^2}{3}} \right), \\
 H(r) = \frac{1}{6r^3} + \left( { - \frac{3}{16\varepsilon ^4} +
\frac{3}{16\varepsilon ^3} + \frac{3}{8\varepsilon ^2} -
\frac{1}{4\varepsilon } + \frac{1}{6} + \frac{1}{12}\varepsilon }
\right)\frac{1}{r^2} + \frac{(2\varepsilon ^2 + 3)(2\varepsilon -
1)}{48\varepsilon ^2r}. \\
 \end{array}
\]

The inside solution is next pairs

\[
\begin{array}{l}
 \chi _3 = z - 0.009z^5 - ... + \log z(0.02z^5 + ...) + b(0.1667z^3 - ...) +
b\log z(...) + ..., \\
 \varphi _3 = 0.01298z^5 + 0.01202z^7 + ... - \log z(0.2828z^5 + ...) +
b(0.0113z^5 + ...) + ...; \\
 \end{array}
\]

\[
\begin{array}{l}
 - \chi _4 = 0.07071z^5 + 0.002405z^7 + ... + b(0.004z^7 + ...), \\
 \varphi _4 = z^3 + 0.00714z^5 + ... + b(0.0714z^5 + ...) + ...; \\
 \end{array}
\]

where

\[
z = \sqrt {B_0 } r;
\]

\[
b = \frac{A_0 }{B_0 } = \frac{\frac{1}{3}a + r_0 \varepsilon ^2e^{r_0
}}{a\left( {\frac{1}{3} + \frac{1}{r_0 } + \frac{1}{r_0^2 }} \right)};
\]

$r_{0}$=0.4fm.

W. Rarita and J. Schwinger obtain the following differential
equations for the $^{3}$S$_{1}$ and $^{3}$D$_{1}$ radial deuteron
wave functions \cite{Rarita1941}

\[
\begin{array}{l}
 \frac{d^2u}{dr^2} + \frac{M}{\hbar ^2}\left[ {E + J} \right]u = - 2^{3 /
2}\gamma \frac{M}{\hbar ^2}Jw, \\
 \frac{d^2w}{dr^2} - \frac{6w}{r^2} + \frac{M}{\hbar ^2}\left[ {E + (1 -
2\gamma )J} \right]u = - 2^{3 / 2}\gamma \frac{M}{\hbar ^2}Ju. \\
 \end{array}
\]

Outside the range of interaction these coupled equations are readily
integrable. The result of such decisions

\[
\begin{array}{l}
 u(r > r_0 ) = Ae^{ - \alpha (r - r_0 )}, \\
 w(r > r_0 ) = Be^{ - \alpha (r - r_0 )}\left( {1 + \frac{3}{\alpha r} +
\frac{3}{(\alpha r)^2}} \right), \\
 \end{array}
\]

where $\alpha = \sqrt {M\left| {E_0 } \right| / \hbar ^2} $, $\left| {E_0 }
\right| = - E = 2.17$MeV.

At distances less than $r_0$ the differential equations for the
ground state wave function will be written in the following form

\[
\begin{array}{l}
 \left( {\frac{d^2}{dr^2} + \kappa ^2} \right)u(r) = - \lambda ^2w(r), \\
 \left( {\frac{d^2}{dr^2} - \frac{6}{r^2} + \kappa '^2} \right)w(r) = -
\lambda ^2u(r). \\
 \end{array}
\]

Here introduced the next notation

\[
\begin{array}{l}
 \kappa ^2 = \frac{M(V_0 - \left| {E_0 } \right|)}{\hbar ^2}, \\
 \kappa '^2 = \frac{M((1 - 2\gamma )V_0 - \left| {E_0 } \right|)}{\hbar ^2},
\\
 \lambda ^2 = \frac{2^{3 / 2}\gamma MV_0 }{\hbar ^2}. \\
 \end{array}
\]

The procedure adopted was the expansion for deuteron wave functions $u(r)$ and
$w(r)$ in infinite power series

\[
\begin{array}{l}
 u(r) = \sum\limits_0^\infty {A_n x^{n + 1}} + \ln x\sum\limits_0^\infty
{C_n x^{n + 2}} , \\
 w(r) = \sum\limits_0^\infty {B_n x^{n + 3}} + \ln x\sum\limits_0^\infty
{D_n x^{n + 3},} \\
 x = r / r_0 . \\
 \end{array}
\]

The constants $A_{n}$, $B_{n}$, $C_{n}$, $D_{n}$ satisfy the recursion formulas

\[
\begin{array}{l}
 (n + 1)(n + 2)A_{n + 1} + (2n + 3)C_n + (\kappa r_0 )^2A_{n - 1} = -
(\lambda r_0 )^2B_{n - 3} , \\
 (n + 1)(n + 2)C_n + (\kappa r_0 )^2C_{n - 2} = - (\lambda r_0 )^2D_{n - 3}
, \\
 n(n + 5)B_n + (2n + 5)D_n + (\kappa 'r_0 )^2B_{n - 2} = - (\lambda r_0
)^2A_n , \\
 n(n + 5)D_n + (\kappa 'r_0 )^2D_{n - 2} = - (\lambda r_0 )^2C_{n - 1} . \\
 \end{array}
\]

The criterion for continuity of the logarithmic derivatives of function
$u(r)$ and $w(r)$ gives two simple equations

\[
\begin{array}{l}
 \left( {\frac{r_0 }{u}\frac{du}{dr}} \right)_{r = r_0 } = - \alpha r_0 , \\
 \left( {\frac{r_0 }{w}\frac{dw}{dr}} \right)_{r = r_0 } = - \left( {2 +
\frac{(\alpha r_0 )^2(1 + \alpha r_0 )}{(\alpha r_0 )^2 + 3\alpha r_0 + 3}}
\right), \\
 \end{array}
\]

which suffice to amply determine $B_{0}/A_{0}$ and $V_{0}$ for a given choice
of parameters $r_{0 }$ and $\gamma $.

The constants $A$ and $B$ may be derived from the known normalization condition:

\[
\begin{array}{l}
 \int\limits_0^\infty {(u^2 + w^2)dr = } \int\limits_0^\infty {(u^2 + w^2)dr
+ \frac{A^2}{2\alpha } + \frac{B^2}{2\alpha }\left( {1 + \frac{6(1 + \alpha
r_0 )^2}{(\alpha r_0 )^3}} \right) = } 1. \\
 \\
 \end{array}
\]

The final set of constants was calculated as $V_0 / \left| {E_0 } \right| =
6.4$; $\gamma $=0.775; $r_{0}$=2.8*10$^{ - 13}$cm.

Inside the range interaction a general expansion for DWF is
\cite{Rarita1948}

\[
\begin{array}{l}
 u(r) = \sum\limits_i {A_i (\kappa _i r)^{1 / 2}J_{1 / 2} (\kappa _i r)} =
\sum\limits_i {u_i (r)} , \\
 w(r) = \sum\limits_j {B_j (\lambda _j r)^{1 / 2}J_{5 / 2} (\lambda _j r)}
= \sum\limits_j {w_j (r)} , \\
 \end{array}
\]

where $u_{i}$ and $w_{j}$ are the modes in terms of Bessel functions of order
one-half or five-halves. The wave-lengths (\textit{$\kappa $}$_{i}$ and \textit{$\lambda $}$_{j})$ of these modes
are determined by the continuity of the logarithmic derivative.

Also different set of modes for the radial functions were taken as an
exponential times a power series for the interparticle distance:

\[
\begin{array}{l}
 u(r) = \sum\limits_i {A_i r^i\exp ( - \lambda r)} = \sum\limits_i {u_i
(r)} , \\
 w(r) = \sum\limits_j {B_j r^{j + 2}\exp ( - \mu r)} = \sum\limits_j {w_j
(r)} . \\
 \end{array}
\]

The parameters $\lambda $ and $\mu $ for radial DWF are practically fixed by
minimizing the energy.

In Ref.  \cite{guindon1948} it is investigated the radial
dependence of the tensor force in the Deuteron. The find the
solutions of coupled Schrodinger equations for DWF with methods
are similar to the ones used by Rarita and Schwinger. Such ranges
and them regions are considered.

A. Range of tensor force equal to range of ordinary force: $\varepsilon $=1.

For region I $r_0 \ge r \ge 0$ were received solutions

\begin{equation}
\label{eq3}
\begin{array}{l}
 u = \sum\limits_n {(A_n + C_n \ln x)x^n,} \\
 w = \sum\limits_n {(B_n + D_n \ln x)x^n} . \\
 \end{array}
\end{equation}

For region II $\infty \ge r \ge r_0 $ solutions is

\begin{equation}
\label{eq4}
\begin{array}{l}
 u = A\exp \left\{ { - \alpha (r - r_0 )} \right\} + C\exp \left\{ {\alpha
(r - r_0 )} \right\}, \\
 u = B\exp \left\{ { - \alpha (r - r_0 )} \right\}\left[ {1 +
\frac{3}{\alpha r} + \frac{3}{(\alpha r)^2}} \right] + D\exp \left\{ {\alpha
(r - r_0 )} \right\}\left[ {1 - \frac{3}{\alpha r} + \frac{3}{(\alpha r)^2}}
\right]. \\
 \end{array}
\end{equation}

B. Range of tensor force less than range of ordinary force: $\varepsilon
$<1.

For region I: $\varepsilon r_0 \ge r \ge 0$ solutions

\[
\begin{array}{l}
 u = \sum\limits_n {(A_n + C_n \ln y)y^n,} \\
 w = \sum\limits_n {(B_n + D_n \ln y)y^n} . \\
 \end{array}
\]

For region II $r_0 \ge r \ge \varepsilon r_0 $ solutions

\[
\begin{array}{l}
 u = A'\sin (\kappa r) + C'\cos (\kappa r), \\
 w = B'\left[ {\sin (\kappa r) + \frac{3}{\kappa r}\cos (\kappa r) -
\frac{3}{(\kappa r)^2}\sin (\kappa r)} \right] + \\
 + D'\left[ {\cos (\kappa r) - \frac{3}{\kappa r}\sin (\kappa r) -
\frac{3}{(\kappa r)^2}\cos (\kappa r)} \right]. \\
 \end{array}
\]

For region III $\infty \ge r \ge r_0 $ solutions is the same as (\ref{eq4}).

C. Range of tensor force greater than range of ordinary force: $\varepsilon
$>1.

For region I $r_0 \ge r \ge 0$ solutions is the same as (\ref{eq3}).

For region II $\varepsilon r_0 \ge r \ge r_0 $ solutions

\[
\begin{array}{l}
 u = \sum\limits_n {(A'_n + C'_n \ln y)y^n,} \\
 w = \sum\limits_n {(B'_n + D'_n \ln y)y^n} . \\
 \end{array}
\]

For region III $\infty \ge r \ge \varepsilon r_0 $ solutions

\[
\begin{array}{l}
 u = A\exp \left\{ { - \alpha (r - \varepsilon r_0 )} \right\} + C\exp
\left\{ {\alpha (r - \varepsilon r_0 )} \right\}, \\
 u = B\exp \left\{ { - \alpha (r - \varepsilon r_0 )} \right\}\left[ {1 +
\frac{3}{\alpha r} + \frac{3}{(\alpha r)^2}} \right] + D\exp \left\{ {\alpha
(r - \varepsilon r_0 )} \right\}\left[ {1 - \frac{3}{\alpha r} +
\frac{3}{(\alpha r)^2}} \right]. \\
 \end{array}
\]

Pairs of the equations for these areas are specified in work
\cite{guindon1948}. The series coefficients satisfy the recurrence
formulas:

\[
\begin{array}{l}
 (n + 1)(n + 2)A_{n + 2} + (2n + 3)C_{n + 2} + aA_n + cB_n = 0, \\
 (n + 1)(n + 2)C_{n + 2} + aC_n + cD_n = 0, \\
 (n - 1)(n + 4)B_{n + 2} + (2n + 3)D_{n + 2} + bB_n + cA_n = 0, \\
 (n - 1)(n + 4)D_{n + 2} + bD_n + cC_n = 0. \\
 \end{array}
\]

Here it is used following abbreviations

\[
a = (\kappa r_0 )^2;a' = (\alpha r_0 )^2;b = (\kappa 'r_0 )^2;b' = (\alpha
'r_0 )^2;c = (\lambda r_0 )^2;
\]

\[
\begin{array}{l}
 \alpha = \frac{\sqrt {ME_0 } }{\hbar };\alpha ' = \frac{\sqrt {M(E_0 +
2\gamma V_0 )} }{\hbar }; \\
 \kappa = \frac{\sqrt {M(V_0 - E_0 )} }{\hbar };\kappa ' = \frac{\sqrt {M([1
- 2\gamma ]V_0 - E_0 )} }{\hbar }; \\
 \lambda = \frac{\sqrt {2^{3 / 2}\gamma MV_0 } }{\hbar };x = \frac{r}{r_0
};y = \frac{r}{\varepsilon r_0 }. \\
 \end{array}
\]

At the outside of potentials NN interaction $u(x)$ and $w(x)$ have
following form \cite{Matsumoto1954}

\[
\begin{array}{l}
 u(r) = N\exp ( - r / \xi ); \\
 w(r) = N'\exp ( - r / \xi )\left[ {3(\xi / r)^2 + 3(\xi / r) + 1} \right],
\\
 \end{array}
\]

where constant $\xi $ is determined from the binding energy of deuteron. The
coupled equations (\ref{eq2}) have two independent solutions, which satisfy the
boundary its conditions and are denoted by $\psi _1 = (u_1 ,w_1 )$, $\psi _2
= (u_2 ,w_2 )$. Any solution of (\ref{eq2}) is given by

\[
\psi _1 + \alpha \psi _2 = (u_1 + \alpha u_2 ,w_1 + \alpha w_2 ).
\]

For core radius $r_{0}$

\[
\begin{array}{l}
 u_1 (r_0 ) + \alpha u_2 (r_0 ) = 0, \\
 w_1 (r_0 ) + \alpha w_2 (r_0 ) = 0, \\
 \end{array}
\]

therefore, $r_{0}$ is the zero point of determinant $\left|
{{\begin{array}{*{20}c}
 {u_1 (x)} \hfill & {u_2 (x)} \hfill \\
 {w_1 (x)} \hfill & {w_2 (x)} \hfill \\
\end{array} }} \right|$ and a is given by $\alpha = - \frac{u_1 (r_0 )}{u_2
(r_0 )}$.

Static parameters determined $\alpha $ by

\[
\alpha = \frac{ - (bX - B)\pm \sqrt D }{aX - A},
\]

where \textit{A, B, C, a, b, c} are some integrals quadratic of wave functions

\[
a = \int {(u_2^2 + w_2^2 )dr} ;
\quad
b = \int {(u_1 u_2 + w_1 w_2 )dr} ;
\quad
c = \int {(u_1^2 + w_1^2 )dr} ;
\]

\[
D = X^2(b^2-ac) - X(2bB - Ac - aC) + (B^2 - AC).
\]

The assumed potentials confine the physical value $X$ to some limited region.
For example, numerical results are given below with $V_{C}$=-500MeV;
$V_{T}$=-500 or 300MeV.

In the method for the solution of the deuteron problem and its
application to a regular potential were applied such sets trial
functions \cite{Cap1955}

\[
\left\{ {\begin{array}{l}
 u = are^{ - \mu r}, \\
 w = bre^{ - \mu r}, \\
 \end{array}} \right.
\]

\[
\left\{ {\begin{array}{l}
 u = ar^2e^{ - \mu r}, \\
 w = br^3e^{ - \mu r}. \\
 \end{array}} \right.
\]

or

\[
\left\{ {\begin{array}{l}
 u(r) = - 0.822\psi _{30} - 0.3965\psi _{31} - 0.2289\psi _{32} - 0.1172\psi
_{33} - 0.1729\psi _{34} ; \\
 w(r) = 2.25466w_0 + 13.6903\psi _{30} - 9.9299\psi _{31} + 0.7286\psi _{32}
+ 0.4131\psi _{33} + 0.1079\psi _{34} , \\
 \end{array}} \right.
\]

where $w_0 = \frac{1}{\sqrt 2 }\psi _{10} $; $\psi _{3i} $ are
Laguerre functions.

A nucleon-nucleon potential which is a well-defined static limit
of a phenomenological covariant interaction is suggested in paper
\cite{noyes1956}. For this model have used a variational wave
function with the correct behavior at the origin and at infinity:

\[
\begin{array}{l}
 u(r) = e^{ - r} - e^{ - \alpha r}, \\
 w(r) = N\left[ {\left( {\frac{3}{r^2} + \frac{3}{r} + 1} \right)e^{ - r} -
\left( {\frac{3}{r^2} + \frac{3\alpha }{r} + \frac{3\alpha ^2 - 1}{2} +
\frac{\alpha r(\alpha ^2 - 1)}{2}} \right)e^{ - \alpha r}} \right], \\
 \end{array}
\]

where $\alpha $=5 and $N$=0.1 are the approximate values of the
variational parameters.

For normalization $\int\limits_0^\infty {\left( {u^2(r) + w^2(r)}
\right)dr = 1} $ of pion-theoretical deuteron function its record
will be as analytical expression \cite{Iwadare1956}

\[
\begin{array}{l}
 u(r) = 1.039\exp ^{ - 0.32r} - 1.392^{ - 2.360r}; \\
 w(r) = 0.02624\left\{ {1 + \frac{3}{0.328r} + \frac{3}{(0.328r)^2}}
\right\}\exp ^{ - 0.328r} - \frac{1.298}{r^2}\exp ^{ - 0.962r}, \\
 \end{array}
\]

In \cite{Moravcsik1958} are desirable to approximate the
Gartenhaus wave function from the cut-off meson theory
\cite{Gartenhaus1955} by an analytic expression. They can be
usefully in the various integrals for calculates phenomena
involving the deuteron. Three such approximations of varying
degrees of accuracy are specified further.

Approximation 1. The best Hulthen type wave function defined by the such
form

\[
u(r) = C\left( {e^{ - \alpha r} - e^{ - \beta r}} \right).
\]

Its parameters $C$ and $\alpha $ are agree with the asymptotic
behavior of the Gartenhaus S- function, and $\beta $ find from the
normalization of the two functions according to formulas

\[
\int\limits_0^\infty {u^2dr} = 4.025;
\]

\[
\int\limits_0^\infty {w^2dr} = 0.29.
\]

The received values of these parameters: $C$=1.85 or 1.91; $\alpha $=0.232;
$\beta $=1.202.

Approximation 2 and 3. Next even better approximation only as sum of
exponential functions has the forms

\[
u(r) = \left\{ {\begin{array}{l}
 C\left( {1 - e^{ - 1.59r}} \right)\left( {e^{ - 0.232r} - e^{ - 1.59r}}
\right), \\
 C\left( {1 - e^{ - 2.5r}} \right)\left( {1 - e^{ - 1.59r}} \right)\left(
{e^{ - 0.232r} - e^{ - 1.90r}} \right). \\
 \end{array}} \right.
\]

A good approximation to the D function using only exponential functions is
the following:

\[
w(r) = \left\{ {\begin{array}{l}
 0.658r^3,\mbox{ 0 < }r\mbox{ < 0.63 } \\
 2.34r^3e^{ - 2r},\mbox{ 0.63 < }r\mbox{ < 2.1 } \\
 0.147e^{ - 0.256r} + 0.810e^{ - 0.577r},\mbox{ 2.1 < }r\mbox{ < + }\infty
\mbox{ } \\
 \end{array}} \right.
\]

which agrees with the Gartenhaus function for D- state everywhere within 4
percent.

For relativistic DWF (In particular for S- state) the authors
\cite{gourdin1959} find as

\[
\psi (r) = \sum\limits_0^\infty {A_q (r)G_0^{2q} \left( {\frac{\pi }{2}}
\right)} ,
\]

where $G_0^{2q} $ are Gegenbauer polynomial at argument $\pi $/2. The radial
DWFs in coordinate and momentum space are Bessel-Fourier transforms to each
other:

\[
A_q (r) = \frac{( - 1)^q}{(2\pi )^2}\int\limits_0^\infty {A_q (p)\frac{J_{2q
+ 1} (pr)}{pr}p^3dp} .
\]

In work \cite{Sakamoto1959} are investigated the elastic
scattering of high energy neutron by deuteron, using DWFs
calculated making use of the meson theoretical potential:

1) The DWF with the hard core:

\[
u(r) = \left\{ {\begin{array}{l}
 N\left\{ {\exp ( - \alpha (r - r_C ) - \exp ( - \beta (r - r_C )}
\right\},\mbox{ }r \le r_C ; \\
 0,\mbox{ }r \le r_C . \\
 \end{array}} \right.
\]

Here $N^2 = \frac{\alpha \beta (\alpha + \beta )}{2\pi (\alpha - \beta
)^2}$.

2) The DWF without the hard core:

\[
u(r) = N\left( {\exp ( - \alpha r) - \exp ( - \beta r)} \right).
\]

For the deuteron state in work \cite{Matsumoto1960} was considered
the pion-theoretical wave function given in \cite{Iwadare1956}.
Thus

\[
\psi = \frac{1}{\sqrt {4\pi } }\left[ {\frac{u(r)}{r} - \frac{1}{\sqrt 8
}S_{12} \frac{w(r)}{r}} \right]\frac{1}{\sqrt 2 }\left( {\xi _1 \eta _2 -
\eta _1 \xi _2 } \right)\chi _m^l .
\]

The plane wave approximation is the conventional form for purpose:

\[
\psi _f = \frac{1}{\sqrt 2 }\left\{ {\xi _1 \eta _2 \exp [ikr] - \xi _2 \eta
_1 \exp [ - ikr]} \right\}\chi _m^l .
\]

Here $k$ is the relative propagation vector of the nucleons; $\xi _{i}$,
$\eta _{i}$ are the isotopic spin wave functions in a proton and a neutron
states; $\chi _m^l $ is the triplet spin function.

For simplicity of calculation for photodisintegration of the
deuteron in the high energy range, are used the following
analytical form which approximates the deuteron wave function very
well in the outer region \cite{Matsumoto1960}:

\[
\begin{array}{l}
 u(r) = A_S \left[ {e^{ - \alpha r} - e^{ - \beta r}} \right], \\
 w(r) = D_1 e^{ - \alpha _1 r} + D_2 e^{ - \alpha _2 r} + D_3 e^{ - \alpha
_3 r}. \\
 \end{array}
\]

The parameters are chosen as:

\[
\begin{array}{l}
 A_S = 1.039; D_1 = 0.111;\alpha _1 = 0.4; \\
 \alpha = 0.328;D_2 = 0.656;\alpha _2 = 1; \\
 \beta = 1.972;D_3 = - 0.767;\alpha _3 = 2. \\
 \end{array}
\]

The wave function and them parameters reproduce result of
calculations for the deuteron parameters: $P_{D}=$7{\%}$,
Q=$2.6$\cdot $10$^{-27}$cm$^{2}$.

In paper \cite{donnachie1962} the deuteron wave-functions used are
of the Hulthen-Sugawara type \cite{Hulthen1957}

\[
\psi _D (r) = \frac{N}{\sqrt {4\pi } }\left\{ {\frac{u_g (r)}{r} +
\frac{S_{12} }{\sqrt 8 }\frac{w_g (r)}{r}} \right\}\chi _m ,
\]

where

\[
\begin{array}{l}
 u_g (r) = \cos \varepsilon _g \left[ {1 - e^{ - \beta (x = x_C )}}
\right]e^{ = x}; \\
 w_g (r) = \sin \varepsilon _g \left[ {1 - e^{ - \gamma (x - x_C )}}
\right]^2e^{ = x}\left[ {1 + \frac{3(1 - e^{ - \gamma x})}{x} + \frac{3(1 -
e^{ - \gamma x})^2}{x^2}} \right]; \\
 \end{array}
\]

$N^{2}$=7.6579$\times $10$^{ - 12}$cm$^{ - 1}$; $x = \alpha r$; $x_C = \alpha
r_C $; $\alpha $=0.2316fm$^{ - 1}$; $r_{C}$ are hard-core radius. Two values
were select for D- probabilities as

$\beta = 7.961; \gamma \quad = 3.798; $sin$\varepsilon _{g}$=0.02666 for 4{\%}
D- state;

$\beta = 7.451; \gamma \quad = 4.799; $sin$\varepsilon _{g}$=0.02486 for 6{\%}
D- state.

The numerical deuteron wave function using the Yale
nucleon-nucleon potential has been approximated by analytic
expressions \cite{Kottler1964}  that contained only exponential
functions. A first approximation consisted with Hulthen function
for S- wave of the form

\[
u_1 (r) = Ae^{ - \alpha r} - Be^{ - \beta r}.
\]

Value of parameters $A$ and $\alpha $ are determined by the asymptotic
behaviour of the radial wave function, $B$ by the boundary conditions at the
hard core and $\beta $ by the required normalization from the S- state. The
result for these parameters is

\[
A = 1.04965;\alpha = 0.331;B = 2.57955;\beta = 2.900.
\]

An improved approximation to $u(x)$ is obtained with the function

\[
u_2 (r) = \left( {1 + 1.039e^{ - 5r} - 8e^{ - 10.58r}} \right)\left(
{1.0459e^{ - 0.331r} - 2.5702e^{ - 2.9r}} \right).
\]

Fit the Yale D- state data were received a suitable approximation with a
function of the form

\[
w(r) = \left\{ {\begin{array}{l}
 Ae^{ - \alpha r} - Be^{ - \beta r},\mbox{ }0.35 \le r \le 3.416; \\
 Ce^{ - \gamma r} + De^{ - \delta r},\mbox{ }3.416 \le r. \\
 \end{array}} \right.
\]

The values of the constants are

\[
\begin{array}{l}
 A = 0.46354;\alpha = 0.6636;B = 0.24479;\beta = 5.4183; \\
 C = 0.13436;\gamma = 0.417;D = 0.85599;\delta = 1.1703. \\
 \end{array}
\]

For the Schrodinger equations for the deuteron radial wave
functions are look for a solution of this equation having the
following form \cite{Bialkowski1963}

\[
\left( {\begin{array}{l}
 u(r) \\
 w(r) \\
 \end{array}} \right) = a\left( {\begin{array}{l}
 f_1 (r) \\
 g_1 (r) \\
 \end{array}} \right)\exp [ - \kappa r] + b\left( {\begin{array}{l}
 f_2 (r) \\
 g_2 (r) \\
 \end{array}} \right)\exp [ - \kappa r].
\]

For the deuteron wave function in both S and D states is constructed
following Martin's method. He allows to written down the analytical
solutions as

\[
\begin{array}{l}
 u(r) = Ae^{ - \kappa r}\left[ {1 + \int\limits_1^2 {e^{ - \alpha r}\rho ^ +
(\alpha )d\alpha } + H\int\limits_1^2 {e^{ - \alpha r}\rho ^ - (\alpha
)d\alpha } } \right], \\
 w(r) = Ae^{ - \kappa r}\left[ {H + \int\limits_1^2 {e^{ - \alpha r}\sigma ^
+ (\alpha )d\alpha } + H\int\limits_0^2 {e^{ - \alpha r}\sigma ^ - (\alpha
)d\alpha } } \right], \\
 \end{array}
\]

where $A = a + b;H = \frac{a - b}{a + b};\rho ^\pm = \frac{1}{2}\left( {\rho
_1 \pm \rho _2 } \right);\sigma ^\pm = \frac{1}{2}\left( {\sigma _1 \pm
\sigma _2 } \right).$ In a Martin's method it was considered that

\[
\begin{array}{l}
 f_\lambda = 1 + \int\limits_0^\infty {\rho _\lambda (\alpha )e^{ - \alpha
r}d\alpha } ; \\
 g_\lambda = \eta _\lambda + \int\limits_0^\infty {\sigma _\lambda (\alpha
)e^{ - \alpha r}d\alpha } ; \\
 \end{array}
\]

are solutions of modified equations Schrodinger

\[
\begin{array}{l}
 f_\lambda ^{''} - 2\kappa f_\lambda ^{'} - U_{C} f_\lambda = U_{T} g_\lambda , \\
 g_\lambda ^{''} - 2\kappa g_\lambda ^{'} - (6/r^2 + U_{m})g_\lambda = U_{T} f_\lambda . \\
 \end{array}
\]

Are considered the ``inner'' part of the interaction in the wave functions
themselves by adding two terms for the two-pion exchange and the repulsive
nucleon core. For couple of functions $u(r)$ and $w(r)$ by solutions will be the
following form as (it dearly fixes the normalization of the functions):

\[
\begin{array}{l}
 u(r) = e^{ - \kappa r}\left[ {1 + \int\limits_1^2 {e^{ - \alpha r}\rho ^ +
(\alpha )d\alpha } + H\int\limits_1^2 {e^{ - \alpha r}\rho ^ - (\alpha
)d\alpha } + \gamma _1 e^{ - \xi _1 r} + \gamma _2 e^{ - \xi _2 r}} \right],
\\
 w(r) = e^{ - \kappa r}\left[ {H + \int\limits_1^2 {e^{ - \alpha r}\sigma ^
+ (\alpha )d\alpha } + H\int\limits_0^2 {e^{ - \alpha r}\sigma ^ - (\alpha
)d\alpha } + \gamma _3 e^{ - \xi _1 r} + \gamma _4 e^{ - \xi _2 r}} \right],
\\
 \end{array}
\]

where $H$, $\gamma _{i}$, $\xi _{i}$ are parameters to be fixed. This
representation for DWFs with tensor forces.

In paper \cite{bialkowski1964} was assumed that the true wave
function is a sum of the ``outer'' part found from the welt-known
OPE potential, and an ``inner'' part. The ``outer'' part more
slowly than ``inner'' part vanishes exponentially with an exponent
between one and two pion masses. Bialkowski \cite{Bialkowski1963}
have proposed the wave function of the form

\[
\left\{ {\begin{array}{l}
 u(r) = u_{outer} + u_{inner} , \\
 w(r) = w_{outer} + w_{inner} , \\
 \end{array}} \right.
\]

\[
\begin{array}{l}
 u_{outer} = Ae^{ - \kappa r}\left[ {1 + \int {\frac{\rho ^ +
(\alpha )e^{ - \alpha r}dr}{\alpha (\alpha + 2\kappa )} + H\int
{\frac{\rho ^ - (\alpha )e^{ - \alpha r}dr}{\alpha (\alpha +
2\kappa )}} } }
\right], \\
 w_{outer} = Ae^{ - \kappa r}\left[ {H + \int {\frac{\sigma ^ +
(\alpha )e^{ - \alpha r}dr}{\alpha (\alpha + 2\kappa )} + H\int
{\frac{\sigma ^ - (\alpha )e^{ - \alpha r}dr}{\alpha (\alpha +
2\kappa )}} }
} \right], \\
 \end{array}
\]

\[
\left\{ {\begin{array}{l}
 u_{inner} = Ae^{ - \kappa r}\left[ {\gamma _1 e^{ - \xi _1 r} + \gamma _2
e^{ - \xi _2 r}} \right], \\
 w_{inner} = Ae^{ - \kappa r}\left[ {\gamma _3 e^{ - \xi _1 r} + \gamma _4
e^{ - \xi _2 r}} \right]. \\
 \end{array}} \right.
\]

Except these forms, are also such forms for ``inner'' part DWF
\cite{bialkowski1964} as

\[
\left\{ {\begin{array}{l}
 u_{inner} = Ae^{ - \kappa r}\left[ {\gamma _1 e^{ - \xi _1 r} + \gamma _2
e^{ - \xi _2 r}} \right], \\
 w_{inner} = Ae^{ - \kappa r}\left[ {\gamma _3 + \gamma _4 } \right]e^{ -
\xi _2 r}; \\
 \end{array}} \right.
\]

\[
\left\{ {\begin{array}{l}
 u_{inner} = Ae^{ - \kappa r}\left[ {\gamma _1 e^{ - \xi _1 r} - \gamma _2
e^{ - \xi _2 r}} \right], \\
 w_{inner} = Ae^{ - \kappa r}\left[ {\gamma _3 e^{ - \xi _3 r} - \gamma _4
e^{ - \xi _2 r}} \right]. \\
 \end{array}} \right.
\]

In the work \cite{McGee1966} authors have approximated the
coordinate space wave functions by a sum of exponentials or Hankel
functions. The deuteron S state can then be viewed as an extension
of the known Hulthen wave function. The wave functions in
coordinate space have the form

\[
\begin{array}{l}
 u(r) = N\left( {e^{ - \alpha r} + \sum\limits_{j = 1}^n {C_j e^{ -
\varepsilon _j r}} } \right), \\
 w(r) = \rho N\left( {\alpha rh_2 (i\alpha r) + \sum\limits_{j - 1}^n {C_j^
/ \varepsilon _j^ / rh_2 (i\varepsilon _j^ / r)} } \right), \\
 \end{array}
\]

where $h_{2}$ is the spherical Hankel function $xh_2 (ix) = e^{ - x}\left[ {1
+ 3 / x + 3 / x^2} \right]$; $\alpha = \sqrt {M\varepsilon } $ is given by
the deuteron binding energy $\varepsilon $. Fitted pole positions and
residues are denoted by $\varepsilon _{j}, C_{j}$. Coefficient $N$ is
normalization for wave function in terms of the deuteron effective range
$\rho $

\[
N^2 = \frac{2\alpha }{1 - \alpha \rho ( - \varepsilon , - \varepsilon )}.
\]

The calculated values of parameters were provided as $\alpha $=0.2338fm$^{ -
1}$; $N$=0.8896fm$^{ - 1 / 2}$; $\rho $=0.0269.

The deuteron wave function may be expanded \cite{elliott1968} in
the complete set of relative oscillator functions $\phi _{nl} $
($s$=1; $j$=1; $l$=0 or 2)

\[
\psi = \sum\limits_{\begin{array}{l}
 n = 0 \\
 \mbox{ }l \\
 \end{array}}^\infty {\alpha _{nl} \phi _{nl} } ,
\]

where \cite{elliott19680}:

\[
\phi _{nl} (r_i ,b) = \sqrt {\frac{2\Gamma \left( {n + l + \frac{3}{2}}
\right)}{b^3n!}} \frac{r_i^l \exp \left( { - \frac{r_i^2 }{2b^2}}
\right)}{b^l\Gamma \left( {l + \frac{3}{2}} \right)}F\left( { - n\left| {l +
\frac{3}{2}} \right|\frac{r_i^2 }{b^2}} \right).
\]

5. \textbf{Analytical forms of DWF in the years 1970-1999}

Yamaguchi's separable tensor potential generates a deuteron wave
function in momentum space. Fourier transformation produces wave
function in coordinate space \cite{Burnap1970}

\[
\begin{array}{l}
 u(r) = e^{ - \alpha r} - e^{ - \beta r}, \\
 w(r) = \eta \left[ {\left( {1 + \frac{3}{\alpha r} + \frac{3}{\alpha
^2r^2}} \right)e^{ - \alpha r} + } \right. \\
 + \left. {\left( {\frac{(\alpha ^2 - \gamma ^2)(\gamma r + 1)}{2\alpha ^2}
- \frac{\gamma ^2}{\alpha ^2} - \frac{3\gamma }{\alpha ^2r} -
\frac{3}{\alpha ^2r^2}} \right)e^{ - \gamma r}} \right], \\
 \end{array}
\]

where the asymptotic ratio of D to S wave

\[
\eta = \mathop {\lim }\limits_{r \to \infty } \left[ {\frac{w(r)}{u(r)}}
\right] = \frac{\alpha ^2(\beta ^2 - \alpha ^2)t}{(\gamma ^2 - \alpha
^2)^2}
\]

Function $w(r)$ is proportional to $r^{2}$:

\[
\mathop {\lim }\limits_{r \to 0} w(r) = \frac{\eta (\gamma ^2 - \alpha
^2)^2}{8\alpha ^2}r^2.
\]

Using function $u(r)$ and $w(r)$ it is possible to find the central potential
$V_{C}(r)$ and the tensor potential $V_{T}(r)$. For this reason Burnap et all. solve
the coupled equations for radial DWF. In the result is written down the
local potentials corresponding to Yamaguchi's form factors as

\[
V_C = \frac{ - \hbar ^2(\beta ^2 - \alpha ^2)}{M}\left[ { - \frac{wt(\gamma
r + 1)}{2u}} \right.e^{ - \gamma r}\left. { + \frac{1 - w / \sqrt 2 }{u}e^{
- \beta r}} \right]\left( {u - \frac{w}{\sqrt 2 } - \frac{w^2}{u}} \right)^{
- 1},
\]

\[
V_T = \frac{ - \sqrt 8 \hbar ^2}{M}\left[ {\frac{(\gamma ^2 - \alpha
^2)^2}{2\alpha ^2}\eta (\gamma r + 1)e^{ - \gamma r}} \right.\left. { -
\frac{w(\beta ^2 - \alpha ^2)}{u}e^{ - \beta r}} \right]\left( {u -
\frac{w}{\sqrt 2 } - \frac{w^2}{u}} \right)^{ - 1}.
\]

Parameters $\beta $, $\gamma $, $t$ are definite in
\cite{Burnap1970}, thus $\alpha $=0.2316fm$^{ - 1}$.

Humberston and Wallace offered some series of analytic
approximations \cite{Humberston1970} to the deuteron wave function
for Hamada-Johnston potential. The solution for coupled equations
for the radial components DWF must satisfy the boundary conditions

\[
\begin{array}{l}
 u(x_0 ) = 0,\mbox{ }u(x) \approx e^{ - \kappa r}, \\
 w(x_0 ) = 0,\mbox{ }w(x) \approx e^{ - \kappa r}\left( {1 + \frac{3}{\kappa
r} + \frac{3}{(\kappa r)^2}} \right), \\
 \end{array}
\]

where $x_{0}=$0.343fm is the hard-core radius.

Equations for the radial components of the S- and D- state wave functions
was then transformed to

\[
\begin{array}{l}
 \left\{ {\frac{d^2}{dy^2} + \frac{2}{y}\frac{d}{dy} - \frac{\kappa ^2}{y^4}
- A(y)} \right\}\bar {u}(y) - B(y)\bar {w}(y) = 0, \\
 \left\{ {\frac{d^2}{dy^2} + \frac{2}{y}\frac{d}{dy} - \frac{6}{y^2} -
\frac{\kappa ^2}{y^4} - C(y)} \right\}\bar {w}(y) - B(y)\bar {u}(y) = 0, \\
 \end{array}
\]

where $y = 1 / r;\bar {u}(y) = u(r);$

\[
\begin{array}{l}
 A(y) = U_C (r) / y^4;B(y) = 2\sqrt 2 U_T (y) / y^4; \\
 C(y) = \left[ {U_C (r) - 2U_T (r) - 3U_{LS} (r) - 3U_{LL} (r)} \right] /
y^4. \\
 \end{array}
\]

Here $U_{j}(r)$ is components a nucleon-nucleon potential.

Forms of analytic approximations to the solution of coupled equations were
obtained for the modified and unmodified Hamada-Johnston potentials. It was
applied the Rayleigh-Ritz variational method to the deuteron binding energy.
The trial function for the deuteron as

\[
\left\{ {\begin{array}{l}
 u(r) = e^{ - \alpha r}\left( {1 - e^{ - \delta (r - r_0 )}}
\right)\sum\limits_{i = 1}^L {c_i e^{ - (i - 1)\mu r}} = \sum\limits_{i =
1}^L {c_i \phi _i } , \\
 w(r) = e^{ - \alpha r}\left( {1 - e^{ - \rho (r - r_0 )}} \right)\left( {1
+ \frac{3}{\alpha r} + \frac{3}{\alpha ^2r^2}} \right)\sum\limits_{i = 1}^N
{d_i e^{ - (i - 1)\mu r}} = \sum\limits_{i = 1}^N {c_{M + i} \phi _{M + i} }
. \\
 \end{array}} \right.
\]

Here $\hbar ^2\alpha ^2 / M = - E_\alpha $ and $\delta $, $\rho $, $c_{i}$
($i$=1,\ldots ,$L)$, $d_{i}$ ($i$=1,\ldots ,$N)$ are variational parameters.

DWFs for Reid soft-core potential are selected according to
\cite{Vary1973}:

(a) particular Haftel-Tabakin cases \cite{Haftel1971}:

\[
\begin{array}{l}
 u(r) = C_0 e^{ - \alpha _0 r}(1 - \beta _0 r), \\
 w(r) = C_2 re^{ - \alpha _2 r}(1 - \beta _2 r), \\
 \end{array}
\]

(b) "fixed-range" cases:

\[
g(r) = \alpha (1 - p)p^a\left[ {1 - bp^c + (b - 1)p^d} \right];0 \le r \le
e,
\]

where $p = 1 - r / e$. Appropriate parameters and properties of the unitary
transformations are presented as UT8, 13, 18, 22, 23 for case (a) and UT101,
102, 103 for case (b).

The resulting form of the separable potentials \cite{pieper1974}
is

\[
\upsilon = \sum\limits_n {\left| {\upsilon _n } \right\rangle \lambda _n
\left\langle {\upsilon _n } \right|} ;
\]

\[
\left\langle {l,p} \right|\left. {\upsilon _n } \right\rangle =
\sum\limits_m {b_{n,l;m} u_{l,m} (p)} ;
\]

where DWF in momentum space

\[
\begin{array}{l}
 u(p) = \frac{1}{(p^2 + \alpha _m^2 )^2}; \\
 w(p) = \frac{p^2}{(p^2 + \alpha _m^2 )^3}. \\
 \end{array}
\]

The Fourier transforms of DWF in momentum space are

\[
\begin{array}{l}
 u(r) = - \sqrt {\frac{\pi }{8}} \frac{\exp ( - \alpha _m r)}{\alpha _m },
\\
 w(r) = - \frac{1}{8}\sqrt {\frac{\pi }{2}} r\exp ( - \alpha _m r). \\
 \end{array}
\]

To determine the unitary pole approximations for a concrete
potential model, in \cite{afnan1975} are done calculations the
two-nucleon bound state wave functions in momentum space. The
partial wave Schrodinger equation appropriate to S- and D- state
it is written down as

\[
\mathchar'26\mkern-10mu\lambda \left( {\frac{d^2}{dr^2} -
\frac{6}{r^2}\delta _{l2} - k_d^2 } \right)u_l (r) = \sum\limits_{L = 0,2}
{V_{lL} (r)u_L (r)} ,
\]

where $k_d^2 = E_d / \mathchar'26\mkern-10mu\lambda $; $E_{d}$ are deuteron
binding energy. To solves the coupled equations components of the radial
deuteron wave function $u_{l}(r)$, use expressions

\[
\begin{array}{l}
 u_l (r) = 0\mbox{ }for\mbox{ }r < r_c , \\
 u_l (r) = - \sqrt {\frac{\pi }{2}} \mathchar'26\mkern-10mu\lambda
\sum\limits_{j = 1}^N {\alpha _l^j \phi _l^j (r)} \mbox{ }for\mbox{ }r \ge
r_c , \\
 \mbox{ } \\
 \end{array}
\]

where $r_{c}$ are hard-core radius; $\alpha _l^j $ are the
expansion coefficients. The effect of the hard core be
incorporated by the modification

\[
\begin{array}{l}
 \phi _0^j (r) = \exp ( - k_d r) - \eta _0^j \exp ( - a_j r), \\
 \phi _0^j (r) = 2k_d^2 A_{5 / 2} (k_d r) - \eta _2^j \left[ {2a_j^2 A_{5 /
2} (a_j r) - (k_d^2 - a_j^2 )a_j rA_{3 / 2} (a_j r)} \right]. \\
 \end{array}
\]

where

\[
\begin{array}{l}
 A_{5 / 2} (\mu r) = \left( {1 + \frac{3}{\mu r} + \frac{3}{(\mu r)^2}}
\right)e^{ - \mu r}, \\
 A_{3 / 2} (\mu r) = \left( {1 + \frac{1}{\mu r}} \right)e^{ - \mu r}. \\
 \end{array}
\]

Here $a_{j}$ ($j$=1,$N)$ are predetermined ranges chosen between
0.7 and 20.0 fm. This approximation was applied for the group
potentials of different types: hard core (Reid hard core
\cite{Reid1968}, Hamada- Johnston \cite{Hamada1962}, Yale
\cite{Yale1962}), soft core (Reid soft core and Alternate Reid
soft core \cite{Reid1968}), super soft core (Tourreil-Sprung A, B
and C \cite{tourreil1973}) and velocity dependent (Bryan-Scott,
Bryan-Gersten, Stagat \cite{stagat1971}, Riewe, and Green,
Ueda-Green II).

In work \cite{Coester1975} is submitted Baker transformation as

\[
\tilde {u}(r) = \sqrt {\left( {\frac{dR}{dr}} \right)} u(R(r)),
\]

where

\[
R = r + a + 2\beta \ln \left[ {\frac{1 + \sqrt {1 + \rho \exp ( - r / \beta
)} }{1 + \sqrt {1 + S} }} \right],
\]

$a$ are hard-core radius; S is determined by the asymptotic

\[
\mathop {\lim }\limits_{r \to \infty } \left[ {R(r) - r} \right] - 0.
\]

Besides in work \cite{Coester1975} are specified exotic shapes by
DWF UT101 \cite{Vary1973} two DWFs obtained from RSC wave
functions by a unitary transformation designed for lower the
D-state probability

\[
\begin{array}{l}
 \tilde {u}(r) = C(r)u(r) + S(r)w(r), \\
 \tilde {w}(r) = - S(r)u(r) + C(r)w(r), \\
 \end{array}
\]

where

\[
S(r) = A_{tr} \frac{\tanh (r / \gamma )\exp ( - (r - \rho ) / \tau )}{1 +
\exp ( - (r - \rho ) / \tau )},
\]

\[
C(r) = \sqrt {1 - S^2(r)} .
\]

Here parameters chosen are $A_{tr}$=0.4472; $\gamma $=0.02fm; $\tau $=0.02fm;
$\rho $=0.8 or 1.9fm.

Accordant \cite{Adler1977} the Hulthen wave function for S state
DWF

\[
u(r) = N\left( {e^{ - \gamma r} - e^{ - \beta r}} \right),\mbox{ }\beta > >
\gamma ,
\]

where $\gamma = \sqrt {M\varepsilon } = 0.2316$fm$^{ - 1}$; $\beta $ be
determined from the triplet effective range parameter with the value
$r_{0}$=1.75fm as approximately

\[
\beta = \frac{3 - \gamma r_0 + \sqrt {\gamma ^2r_0^2 - 10\gamma r_0 + 9}
}{2r_0 } = 5.98\gamma .
\]

The normalization constant $N$ in terms of the effective range as

\[
N^2 = \frac{2\gamma }{1 - \gamma r_0 } = 0.783.
\]

Wave function for D- state choose explicitly as

\[
w(r) = \eta N\left( {1 - e^{ - \tau r}} \right)^5e^{ - \gamma r}\left( {1 +
\frac{3}{\gamma r} + \frac{3}{\gamma ^2r^2}} \right).
\]

Multiplication is considered the asymptotic by an interpolating factor.

Formulas for calculation of values of the D-state percentage and for the
quadrupole moment will be respectively

\[
P_D = \eta ^2N^2\sum\limits_{n = 1}^4 {\left[ {\frac{a_n }{(n -
1)!}\sum\limits_{q = 0}^{10} {\left( {\begin{array}{l}
 10 \\
 q \\
 \end{array}} \right)( - 1)^{n - q}(2\gamma + q\tau )^{n - 1}\ln (2\gamma +
q\tau )} } \right]} + \eta ^2N^2a_0 \sum\limits_{q = 0}^{10} {\left(
{\begin{array}{l}
 10 \\
 q \\
 \end{array}} \right)\frac{( - 1)^q}{2\gamma + q\tau }} ;
\]

\[
\begin{array}{l}
 Q = \frac{\eta N^2}{\sqrt {50} }\sum\limits_{n = 0}^5 {b_n } \sum\limits_{q
= 0}^5 {\left( {\begin{array}{l}
 5 \\
 q \\
 \end{array}} \right)( - 1)^qn!} \left[ {\frac{1}{(qr + 2\gamma )^{n + 1}} -
\frac{1}{(q\tau + \gamma + \beta )^{n + 1}}} \right] - \\
 - \frac{\eta ^2N^2}{20}\left\{ {\sum\limits_{n = 0}^2 {c_n } \sum\limits_{q
= 0}^{10} {\left( {\begin{array}{l}
 10 \\
 q \\
 \end{array}} \right)\frac{( - 1)^qn!}{(2\gamma + q\tau )^{n + 1}}} +
\sum\limits_{n = 0}^2 {c_n } \sum\limits_{q = 0}^{10} {\left(
{\begin{array}{l}
 10 \\
 q \\
 \end{array}} \right)\frac{( - 1)^{n - q}}{(n - 1)!}\frac{\ln (2\gamma +
q\tau )}{(2\gamma + q\tau )^{1 - n}}} } \right\}, \\
 \end{array}
\]

where

\[
a_n = \left( {1,\frac{6}{\gamma },\frac{15}{\gamma ^2},\frac{18}{\gamma
^3},\frac{9}{\gamma ^4}} \right);b_n = \left( {\frac{3}{\gamma
^2},\frac{3}{\gamma },1} \right);c_n = \left( {\frac{15}{\gamma
^2},\frac{6}{\gamma },1} \right);d_n = \left( {\frac{18}{\gamma
^3},\frac{9}{\gamma ^4}} \right).
\]

The calculated values of parameters: $\tau $=1.09fm$^{ - 1}$; $\eta $=0.025
for $P$=7{\%} and $\tau $=0.83fm$^{ - 1}$; $\eta $=0.029 for $P$=4{\%}.

In \cite{mcgurk1977} DWF modelled on that of the Reid soft-core
potential (RSCP) outside $1.5\lambda _\pi $:

\[
\psi _L (r) = \left\{ {\begin{array}{l}
 \sum\limits_{i = 1}^8 {a_{Li} r^{i - 1}} ,\mbox{ }r < 1.5\lambda _\pi ; \\
 \psi _{L(RSCP)} (r),\mbox{ }r \ge 1.5\lambda _\pi , \\
 \end{array}} \right.
\]

where $\lambda _\pi $ is the pion Compton wavelength. In radial wave
functions five of the coefficients $a_{Li}$ are determined by: 1) continuity
of DWF together with its first and second derivatives of the RSCP at
$1.5\lambda _\pi $; 2) $u$(0)=0; $w$(0)=0; 3) adjusting a D- state percentage
(4.5-6.5{\%}) and the overall normalization as 1.

In Refs. \cite{Allen1978} and \cite{allen1979} it is considered
electron-deuteron tensor polarization and the short range behavior
of the deuteron wave function. Interactions for twelve classes
varying in the core region obtained using form factor for the
unitary transformation

\[
g(r) = \left\{ {\begin{array}{l}
 C(R - r)^\alpha (1 - \beta r),\mbox{ }r \le R; \\
 0,\mbox{ }r > R. \\
 \end{array}} \right.
\]

where $R$=0.7fm; $\alpha $=2.1. The constant $C$ is determined by
the normalization condition. At a choice $\alpha $>2 from that the
transformed DWF will be continuous and continuous it first and
second derivatives at $R$. Calculations are compared for super
soft-core (SSC) potential \cite{tourreil1973}. The tensor
polarization for the recoil deuterons in \textit{ed} scattering
are calculated as

\[
P_e = \frac{2G_0 G_2 + G_2^2 / \sqrt 2 }{G_0^2 + G_2^2 }.
\]

Its values in the range 0.625-0.668.

Lomon-Feshbach, Holinde-Machleidt and four-component relativistic
models were used for research elastic electron-deuteron scattering
at high energy \cite{Arnold1980}.

In coordinate space expansion in Hulthen functions of different range is
presented as

\[
\frac{u(r)}{r} = \sqrt {\frac{\pi }{2}} \sum\limits_i {c_i
\frac{\exp ( - \beta _i r)}{r}} .
\]

If we calculate the nth moment of the coefficients as

\[
M_n = \sum\limits_i {c_i \beta _i^n } ,
\]

then the reduced wave function \textit{u(r)} will go like $r^{n}$
at the origin.

The normalized solutions of the Schrddinger equation select in
\cite{klarsfeld1981} as

\[
\begin{array}{l}
 u(r) = N\left[ {u_1 (r) + \eta u_2 (r)} \right], \\
 w(r) = N\left[ {w_1 (r) + \eta w_2 (r)} \right], \\
 \end{array}
\]

The experimental values of deuteron observables severely restrict values of
$\eta $. For placing upper and lower bounds for $\eta $ it is used Schwarz's
inequality

\[
U_2 W_2 \ge X^2 + \sqrt {\frac{1}{2}} XW_2 + \frac{1}{8}W_2^2 .
\]

The condition for the existence of a solution is

\[
\Delta (R,\eta ) = Y^2 - 4X^2 - \sqrt 2 XY \ge 0,
\]

where

\[
X = X(R,\eta ) = \sqrt {50} Q - \int\limits_R^\infty {r^2(uw - \sqrt
{\frac{1}{8}} } w^2)dr = V_2 - \sqrt {\frac{1}{8}} W_2 ;
\]

\[
Y = Y(R,\eta ) = 4\left\langle {r^2} \right\rangle - \int\limits_R^\infty
{r^2(u^2 + } w^2)dr = U_2 + W_2 ;
\]

\[
U_n = \int\limits_0^R {r^nu^2dr} ;
\quad
V_n = \int\limits_0^R {r^nuwdr} ;
\quad
W_n = \int\limits_0^R {r^nw^2dr} .
\]

Value$ p_{D}$ it is determined with a condition as

\[
p_D = \int\limits_0^\infty {w^2dr} = W_0 + Z;
\]

\[
p_D > Z + \frac{X^2(1 + sgnX)}{2U_4 };
\]

where $Z = Z(R,\eta ) = \int\limits_R^\infty {w^2dr} $.

In paper \cite{Koike1981} were presented DWFs from Yamaguchi type
form factors with 4{\%} or 7{\%} deuteron D- state probability
(designations YY4 and YY7). Also are obtain a new set T4D-1
(T4D-2) which has the rank-1 (rank-2) separable potential with the
first (second) form factor of T4D.

It should be noted that the most popular, the quoted and used
parametrization of DWF are the analytical forms offered by the
Paris group. Known numerical values of radial DWF in coordinate
representation for the Paris potential can be approximated by
means of convenient decompositions \cite{Lacombe1981} in an such
form:

\begin{equation}
\label{eq5}
\left\{ {\begin{array}{l}
 u\left( r \right) = \sum\limits_{j = 1}^N {C_j \exp \left( { - m_j r}
\right),} \\
 w\left( r \right) = \sum\limits_{j = 1}^N {D_j \exp \left( { - m_j r}
\right)\left[ {1 + \frac{3}{m_j r} + \frac{3}{\left( {m_j r} \right)^2}}
\right],} \\
 \end{array}} \right.
\end{equation}

where $N$=13; $m_j = \beta + (j - 1)m_0 $; $\beta = \sqrt {ME_d } $;
$m_{0}$=0.9fm$^{ - 1}$. $M$ is nucleon mass, $E_{d}$ is binding energy of
deuteron. The boundary conditions as $r \to 0$

\[
u\left( r \right) \to r,
\quad
w\left( r \right) \to r^3.
\]

The asymptotics behavior of the deuteron wave functions for large values of
$r \to \infty $ are

\[
\begin{array}{l}
 u(r) = A_S \exp ( - \beta r), \\
 w(r) = A_D \exp ( - \beta r)\left[ {1 + \frac{3}{\beta r} + \frac{3}{(\beta
r)^2}} \right], \\
 \end{array}
\]

The last coefficients of an analytical form were determined by formulas

\begin{equation}
\label{eq6}
\left\{ {\begin{array}{l}
 C_n = - \sum\limits_{j = 1}^{n - 1} {C_j } ; \\
 D_{n - 2} = \frac{m_{n - 2}^2 }{\left( {m_n^2 - m_{n - 2}^2 } \right)\left(
{m_{n - 1}^2 - m_{n - 2}^2 } \right)}\left[ { - m_{n - 1}^2 m_n^2
\sum\limits_{j = 1}^{n - 3} {\frac{D_j }{m_j^2 } + \left( {m_{n - 1}^2 +
m_n^2 } \right)\sum\limits_{j = 1}^{n - 3} {D_j - \sum\limits_{j = 1}^{n -
3} {D_j m_j^2 } } } } \right] \\
 \end{array}} \right.
\end{equation}

and taking into account conditions

\begin{equation}
\label{eq7}
\sum\limits_{j = 1}^N {C_j } = 0;
\quad
\sum\limits_{j = 1}^N {D_j } = \sum\limits_{j = 1}^N {D_j m_j^2 } =
\sum\limits_{j = 1}^N {\frac{D_j }{m_j^2 }} = 0.
\end{equation}

The accuracy of parametrization is characterized by the values

\[
I_S = \left( {\int\limits_0^\infty {\left[ {u(r) - u_{aprox} (r)}
\right]^2dr} } \right)^{1 / 2},
\]

\[
I_D = \left( {\int\limits_0^\infty {\left[ {w(r) - w_{aprox} (r)}
\right]^2dr} } \right)^{1 / 2}.
\]

Model radial DWF \cite{Certov1987} according to parametrization
(\ref{eq5}) \cite{Lacombe1981} are constructed to facilitate the
exploration of dependencies on the percentage D state and on the
small-, medium-, and large-distance parts of DWF. Parametrization
\cite{Lacombe1981} was also used for approximation of DWF received
for the following potentials: of the (energy-dependent) full model
and from the (energy-independent) relativistic momentum space
OBEPQ \cite{Machleidt1987}, OBEP model A, B, C
\cite{Machleidt1989}, OBEPR, OBEPR(A) and OBEPR(B)
\cite{Levchuk1995} an $N$=11.

Theoretical values for the central and tensor components of the
polarizability are presented in Ref. \cite{Lopes1983}. The are
sums of bilinear combinations of integrals of the form

\[
I(J;L) = \int {r^3u_L (r)f_J (K,r)dr} ,
\]

where $f_{J}$ and $u_{L}$ is the radial wave function of the P wave continuum
and deuteron respectively. The presence of the $r^{3}$ factor strongly
suggests that the long range area of ground state DWF

\[
\begin{array}{l}
 u(r) = A_S \frac{e^{ - \gamma r}}{r}; \\
 u(r) = \eta A_S \frac{e^{ - \gamma r}}{r}\left[ {1 + \frac{3}{\gamma r} +
\frac{3}{(\gamma r)^2}} \right] \\
 \end{array}
\]

will be of value in determining the $I(J;L)$, and hence the calculated
polarizability. Further is investigated the extent to which $\alpha $ and r
are in fact determined by $A_{S}$ and $\eta $.

DWF \cite{deloff1984} must belong to the area of the Hilbert space
orthogonal to the trivial solution. therefore the
orthogonalization is straightforward for the Paris wave function
$u(r)$ and $w(r)$

\[
\begin{array}{l}
 \tilde {u}(r) = \frac{u(r) - C\Phi _0 (r)}{\sqrt {1 - C^2} }; \\
 \tilde {w}(r) = \frac{w(r)}{\sqrt {1 - C^2} }. \\
 \end{array}
\]

Here $b$ is the oscillator width parameters; constant $C$ equal to the product
$\left\langle u \right.\left| {\Phi _0 } \right\rangle $; $\Phi _0 (r)$ is
the eigenfunction of the norm kernel calculated in oscillator basis

\[
\Phi _0 (r) = \left[ {\frac{2}{\pi }\left( {\frac{3}{b^2}} \right)^3}
\right]^{1 / 4}r\exp \left\{ { - \frac{3r^2}{4b^2}} \right\}.
\]

The modified DWF takes the form

\[
\begin{array}{l}
 \tilde {u}(r) = Au(R)\frac{\frac{1}{3}\sin \alpha \Phi _0 (r) -
\frac{1}{5}\cos \alpha \Phi _2 (r)}{\frac{1}{3}\sin \alpha \Phi _0 (R) -
\frac{1}{5}\cos \alpha \Phi _2 (R)};r \le R; \\
 \tilde {u}(r) = Au(r);r \ge R; \\
 \tilde {w}(r) = Aw(r), \\
 \end{array}
\]

where $R$ is certain radius, when for $r$<$R$ the wave function is determined by six
quarks dynamics; $\Phi _0 (r)$ and $\Phi _2 (r)$ are the oscillator wave
functions for the ground state and the level with two excitation quanta. The
ratio between them is such

\[
\Phi _2 (r) = \Phi _0 (r)\sqrt {\frac{3}{2}} \left( {1 - \frac{r^2}{b^2}}
\right).
\]

In paper \cite{klarsfeld1984} a method has been obtained which
determines as whether or not the long-range part for potential
model of a two-body is consistent with measured deuteron
properties and independent of the short-range behaviour. For the
determination outer part of the deuteron wave function was to
construct two independent solutions of the coupled Schrodinger
equations $\left( {\begin{array}{l}
 u_1 \\
 w_1 \\
 \end{array}} \right)$ and $\left( {\begin{array}{l}
 u_2 \\
 w_2 \\
 \end{array}} \right)$ in the region $r \ge R$. Further are used the
asymptotic boundary conditions as

\[
\left( {\begin{array}{l}
 u_1 \\
 w_1 \\
 \end{array}} \right) \to \left( {\begin{array}{l}
 e^{ = x} \\
 \eta _0 xk_2 (x) \\
 \end{array}} \right);
\quad
\left( {\begin{array}{l}
 u_2 \\
 w_2 \\
 \end{array}} \right) \to \left( {\begin{array}{l}
 0 \\
 xk_2 (x) \\
 \end{array}} \right),
\]

where $x = \alpha r$, $xk_2 (x) = e^{ = x}\left( {1 + 3 / x + 3 / x^2}
\right)$. The first solution corresponds to $\eta =\eta _{0}$ and for
other solution $\eta $ take a linear combination:

\[
\left( {\begin{array}{l}
 u \\
 w \\
 \end{array}} \right)_\eta = \left( {\begin{array}{l}
 u_1 \\
 w_1 \\
 \end{array}} \right) + (\eta - \eta _{_0 } )\left( {\begin{array}{l}
 u_2 \\
 w_2 \\
 \end{array}} \right).
\]

In \cite{locher1984} is specified fit the electromagnetic form
factors of the deuteron on the basis of nonrelativistic wave
functions

\[
\begin{array}{l}
 u(r) = N\left[ {e^{ - \alpha r} - \sum\limits_i {c_C^i e^{ - \beta _S^i
r}} } \right]; \\
 w(r) = \rho N\left[ {\alpha rh_2 (i\alpha r) - \sum\limits_i {\left(
{\frac{\beta _D^i }{\alpha }} \right)^2c_D^i \beta _D^i rh_2 (i\beta _D^i
r)} } \right], \\
 \end{array}
\]

where $xh_2 (ix) = \left[ {1 + 3 / x + 3 / x^2} \right]\exp ( - x)$.
Asymptotics at $r \to 0$ for the S and D state will be as

$u(r) = r$or $\sum\limits_i {c_S^i } = 1$;

$u(r) = r^3$or $\left\{ {\begin{array}{l}
 \sum\limits_i {\beta _S^i c_S^i = \alpha } ; \\
 \sum\limits_i {\left( {\beta _S^i } \right)^2c_S^i = \alpha ^2} ; \\
 \end{array}} \right.$

$w(r) = r^3$or $\left\{ {\begin{array}{l}
 \sum\limits_i {c_D^i } = 1; \\
 \sum\limits_i {\left( {\beta _D^i } \right)^2c_D^i } = \alpha ^2; \\
 \sum\limits_i {\left( {\beta _D^i } \right)^4c_D^i } = \alpha ^4. \\
 \end{array}} \right.$

For separable potentials with and without tensor force are
presented calculation of deuteron form factors \cite{Mehdi1984},
which are expressed through radial DWF in configuration space. The
expressions Mehdi-Gupta parametrization for the radial DWF are:

\[
\left\{ {\begin{array}{l}
 u(r) = A\left( {e^{ - \alpha r} - e^{ - \beta r}} \right) + Bre^{ - \beta
r}, \\
 w(r) = C\left[ {\frac{\alpha ^2}{3}\left( {e^{ - \alpha r} - e^{ - \gamma
r}} \right) - \frac{\gamma (\gamma ^2 - \alpha ^2)}{6}re^{ - \gamma r} + }
\right. \\
 \left. { + \left( {\frac{1}{r^2} + \frac{\alpha }{r}} \right)e^{ - \alpha
r} - \left( {\frac{1}{r^2} + \frac{\gamma }{r} + \frac{\gamma ^2 - \alpha
^2}{2}} \right)e^{ - \gamma r}} \right], \\
 \end{array}} \right.,
\]

where $C = \frac{3\sqrt 2 \pi Nt}{(\gamma ^2 - \alpha ^2)^2}$. Coefficients
$A$ and $B$ for shape-1:

\[
A = \frac{\sqrt 2 \pi N}{\beta ^2 - \alpha ^2};
\quad
B = 0;
\]

and for shape-2:

\[
A = \frac{\sqrt 2 \pi N}{(\beta ^2 - \alpha ^2)^2};
\quad
B = - \frac{\pi N}{\sqrt 2 \beta (\beta ^2 - \alpha ^2)}.
\]

The two-body parameters represented as ratios $\beta $/$\alpha $ and $\gamma
$/$\alpha $. The D- state probability $P_{D}$ is given by

\[
P_D = \frac{N^2\pi ^2t^2(5\alpha + \gamma )}{8\gamma (1 + \gamma )^5}.
\]

The following parameterization of DWF for realistic superdeep
local NN- potential (Moscow) was written down as gaussian
expansions \cite{Krasnopolsky1985}

\[
\left\{ {\begin{array}{l}
 u(r) = r\sum\limits_{i = 1}^{N_S } {a_i \exp ( - \alpha _i r^2),} \\
 w(r) = r^3\sum\limits_{i = 1}^{N_D } {b_i \exp ( - \beta _i r^2),} \\
 \end{array}} \right.
\]

where

\[
\begin{array}{l}
 \alpha _i = \frac{\alpha _0 }{41.47}tg^{7 / 2}\left[ {\frac{\pi \left( {2i
- 1} \right)}{4N_S }} \right], \\
 \beta _i = \frac{\beta _0 }{41.47}tg^{7 / 2}\left[ {\frac{\pi \left( {2i -
1} \right)}{4N_D }} \right], \\
 \end{array}
\]

$\alpha _0 = 31,9;\beta _0 = 164;N_{S}=N_{D}$=30.

In \cite{Kalashnikova1986} are considered quark compound bag (QCB)
and six quark bag models and are inquire into the values of
$P_{OCB}$ and $P_{6q}$ predicted by the QCB model. For
illustration the method first consider a "toy model" of the S-
wave deuteron without the \textit{NN} interaction

\[
u(r) = N\left\{ {\begin{array}{l}
 - \gamma _1 sh(\kappa r) + \gamma _2 \sin (\beta r),r \le b, \\
 \exp ( - \kappa r),r \ge b. \\
 \end{array}} \right.
\]

Calculated value were $P_{QCB}$=0.9{\%}; $P_{6q}$=17{\%}. The general
expression for the deuteron wave function in the QCB model it will be
written down as

\[
u_l (r) = N\left\{ {\begin{array}{l}
 b_1 u_l^{(\ref{eq1})} (r) + b_2 u_l^{(\ref{eq2})} (r),r \le b; \\
 u_l^{ext} (r),r \ge b, \\
 \end{array}} \right.
\]

where $N$ is the normalization factor, $u_l^{ext} $ are DWF derived from the
external potential, and $u_l^{(\ref{eq1})} $, $u_l^{(\ref{eq2})} $ are the two linear
independent solutions of Schrodinger equation in the inner region. The
constants $b_{1}$ and $b_{2}$ are defined from the atching condition of the
internal and external wave functions at $r=b$. Thus are established the upper
limit on 1.2fm$ \le b \le $1.6fm ($P_{QCB} \le $1{\%}).

In \cite{oteo1988} are consider a more general case for
\cite{Hulthen1957} and \cite{Adler1977} by including additional
terms as such follows

\[
\left\{ {\begin{array}{l}
 u(r) = A_S (1 - e^{ - \tau r})e^{ - \alpha r}\sum\limits_{i = 0}^n {C_i
\exp ( - \alpha _i r),} \\
 u(r) = \eta A_S (1 - e^{ - \sigma r})^5k_2 (\alpha r)\sum\limits_{i = 0}^m
{D_i \exp ( - \alpha _i r),} \\
 \end{array}} \right.
\]

where \textit{$\alpha $}=0,2315370~fm$^{ - 1}$; \textit{$\tau $}=5\textit{$\alpha $}; \textit{$\sigma $}=1,09~fm$^{ - 1}$; \textit{$\eta $}=0,025; $k_2 (\alpha
r)$ - terms of the spherical Bessel function:

\[
k_2 (\alpha r) = \left( {1 + \frac{3}{\alpha r} + \frac{3}{(\alpha r)^2}}
\right)e^{ - \alpha r}.
\]

Also in \cite{oteo1988} are calculate the simplest
phenomenological realistic deuteron wave function given by
\cite{McGee1966} and \cite{Moravcsik1958}

\[
\left\{ {\begin{array}{l}
 u(r) = A_S (1 - e^{ - \tau r})e^{ - \alpha r}, \\
 u(r) = \eta A_S (1 - e^{ - \sigma r})^5e^{ - \alpha r}\left( {1 +
\frac{3}{\alpha r} + \frac{3}{(\alpha r)^2}} \right). \\
 \end{array}} \right.
\]

Values are obtained for the parameters $\eta $, $\tau $, $\sigma $, when the
indicated values of $A_{S}$ and $r_{d}$ are used as input.

In \cite{grach1990} present a quark compound bag (QCB)
parameterization in r-space. Details of this parameterization are
given in ref. \cite{Kalashnikova1986}. In terms of S- and D- waves
(respectively $l$=0;2) one has

\[
u_l (r) = N\left\{ {\begin{array}{l}
 b_1 u_l^{(\ref{eq1})} (r) + b_2 u_l^{(\ref{eq2})} (r),r \le b; \\
 b_1 u_l^{ext} (r),r > b, \\
 \end{array}} \right.
\]

where $N$ is the normalization factor; $u_{ext}(r)$ are the DWF derived from the
assumed external potential; $u_{ext}(r)$ may be parameterized in terms of Yukawa
functions

\[
\begin{array}{l}
 u_0^{\mbox{ext}} (r) = \sum\limits_{j = 1}^m {C_j \exp ( - m_j r)} ; \\
 u_2^{\mbox{ext}} (r) = \sum\limits_{j = 1}^n {D_j \exp ( - m_j r)\left( {1
+ \frac{3}{m_j r} + \frac{3}{(m_j r)^2}} \right)} . \\
 \end{array}
\]

In \cite{grach1990} are present the QCB model parameters for
$b$=1.2 and 1.35 fm that were selected as representative
solutions.

In Ref. \cite{Mrowczynski1992} are used in calculations the DWF in
the Hulthen form

\[
\phi _d (r) = \sqrt {\frac{\alpha \beta (\alpha + \beta )}{2\pi (\alpha -
\beta )^2}} \frac{e^{ - \alpha r} - e^{ - \beta r}}{r},
\]

where $\alpha $=0.23fm$^{ - 1}$; $\beta $=1.61fm$^{ - 1}$. It is necessary
for receiving the deuteron formation rate

\[
A = \frac{3\pi ^3}{r_0 v\tau }\int\limits_0^\infty {r\left| {\phi _d (r)}
\right|^2\exp \left( { - \frac{r^2}{4r_0^2 }} \right)erfi(ar)dr} ,
\]

\[
a = \frac{v\tau }{2r_0 \sqrt {r_0^2 + v^2\tau ^2} };
\quad
erfi(x) = \frac{2}{\sqrt \pi }\int\limits_0^x {e^{t^2}dt} .
\]

It is also possible to remember also the parameterization of
function received for Moscow \textit{NN }model \cite{kukulin1992}
($N$=24)

\[
\begin{array}{l}
 u\left( r \right) = r\sum\limits_{i = 1}^N {a_i \exp \left( { - \alpha _i
r^2} \right),} \\
 w\left( r \right) = r^3\sum\limits_{i = 1}^N {b_i \exp \left( { - \beta _i
r^2} \right).} \\
 \end{array}
\]

\textbf{6. ``Improved'' analytical forms of DWF}

In some papers the above table of values and coefficients for the
parameterization (\ref{eq5}) \cite{Lacombe1981} and calculated for
him DWF. It is about papers \cite{Certov1987},
\cite{Machleidt1987}, \cite{Machleidt1989}, where there are,
although not insignificant, the obvious knots to DWF near the
origin! In addition, there is an obvious failure to comply with
mandatory conditions for the summation of the coefficients

\[
\sum\limits_{j = 1}^N {C_j } = 0;
\quad
\sum\limits_{j = 1}^N {D_j } = 0.
\]

In the comparative Table 3 shows the results of the summation of the
coefficients of these works.

For my numerical calculations the resulting coefficients are shown Table 4.
The following is the corresponding Fig. 2 where there are knots for WDF.

Table 3. The results of the summation of the coefficients

\begin{table}[htbp]
\begin{tabular}
{|p{140pt}|p{99pt}|p{108pt}|}
\hline
Model&
$\Sigma C_{i}$&
$\Sigma D_{i}$ \\
\hline $\psi _{MM}^{4A} $\cite{Certov1987}& -0.00002&
-0.000055 \\
\hline $\psi _{SL}^{4A} $\cite{Certov1987}& 0.00008&
-0.000325 \\
\hline $\psi _{SL}^{4B} $\cite{Certov1987}& -0.00003&
-0.000565 \\
\hline $\psi _{LS}^{6B} $\cite{Certov1987}& 0.00018&
0.000335 \\
\hline OBEPA \cite{Machleidt1989}& -5.326E-06&
-6.499E-08 \\
\hline OBEPB \cite{Machleidt1989}& -2.471E-06&
1.799E-08 \\
\hline OBEPC \cite{Machleidt1989}& 5.171E-06&
4.710E-07 \\
\hline
\end{tabular}
\label{tab3}
\end{table}

Table 4. Coefficients $D_{i}$ for OBEPC and $\psi _{LS}^{6B}$

\begin{table}[htbp]
\begin{tabular}
{|p{149pt}|p{117pt}|}
\hline
OBEPC&
$\psi _{LS}^{6B} $ \\
\hline
-0.6333000560135323454&
14.8969971094926 \\
\hline
1.53039457858156441734&
-11.2303580952213 \\
\hline
-0.9017034635680319384&
2.835625985728753 \\
\hline
\end{tabular}
\label{tab4}
\end{table}

\pdfximage width 135mm {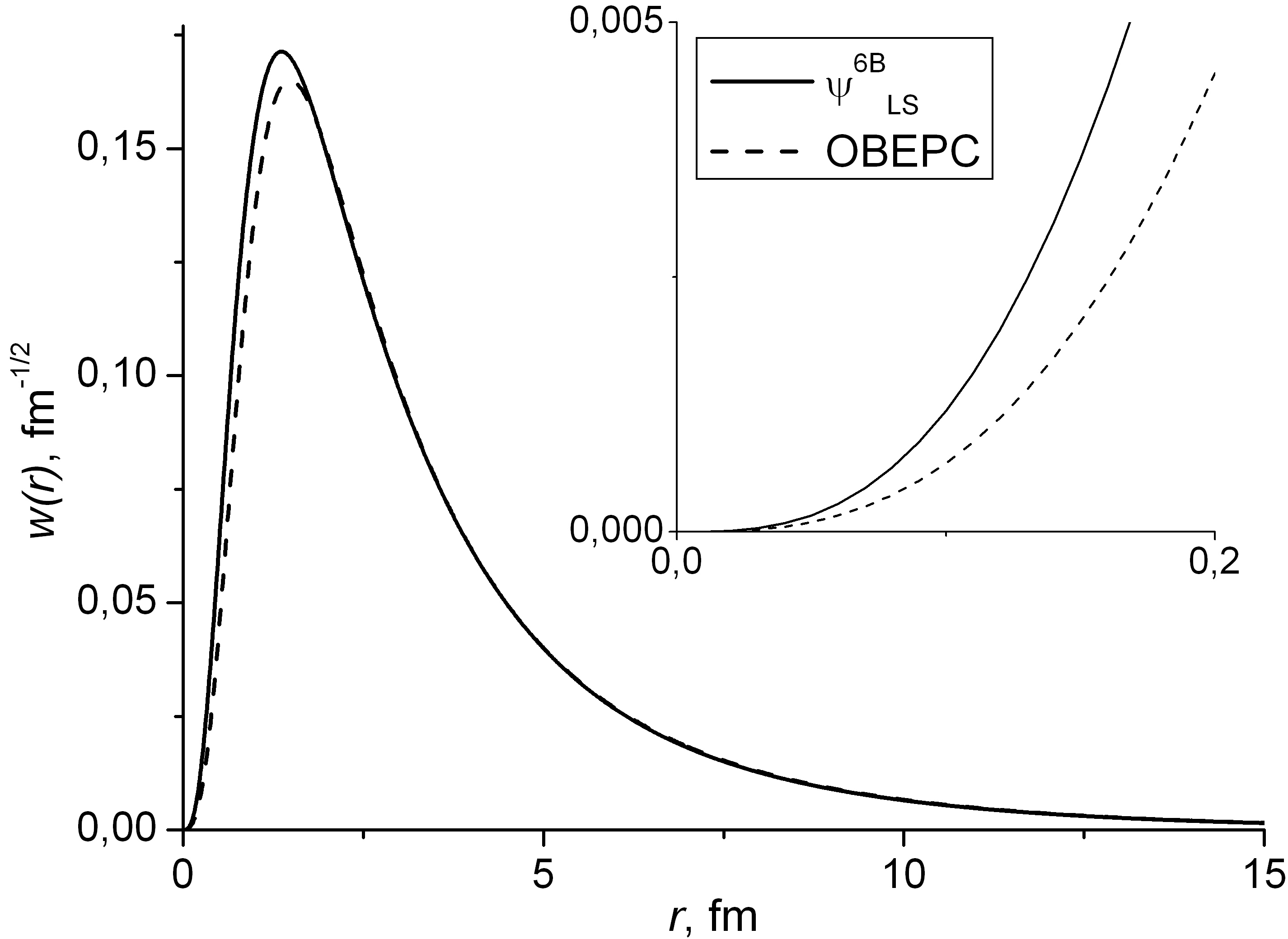}\pdfrefximage\pdflastximage

Fig. 2. ``Improved'' DWF for OBEPC and $\psi _{LS}^{6B}$

So, more accurately calculate the coefficients for the relevant
parameterizations of DWF.

7. \textbf{New analytical forms of DWF}

In 2000-x years are new analytical forms of deuteron wave
function. Except the mentioned parametrization, in literature
there is one more analytical form \cite{Dubovichenko20002} for
DWFs

\begin{equation}
\label{eq8}
\left\{ {\begin{array}{l}
 u(r) = \sum\limits_{i = 1}^N {A_i \exp ( - a_i r^2),} \\
 w(r) = r^2\sum\limits_{i = 1}^N {B_i \exp ( - b_i r^2).} \\
 \end{array}} \right.
\end{equation}

This parametrization was used \cite{Dubovichenko2004} for Nijmegen
potentials groups (NijmI, NijmII, Nijm93 and Reid93). Thus value
$N$=13.

For explanation D- state of deuteron and correct asymptotic
behavior are received nonrelativistic deuteron wave function
\cite{Berezhnoy2005}:

\begin{equation}
\label{eq9}
\begin{array}{l}
 u(r) = \frac{N}{\sqrt {4\pi } }\sum\limits_{k = 1}^{n_u } {C_k \exp ( -
\alpha _k r)} , \\
 u(r) = \frac{N}{\sqrt {4\pi } }\rho \sum\limits_{k = 1}^{n_w } {D_k \exp (
- \beta _k r)} \left( {1 + \frac{3}{\beta _k r} + \frac{3}{(\beta _k r)^2}}
\right), \\
 \end{array}
\end{equation}

\[
N = \sqrt {\sum\limits_{k,j = 1}^{n_u } {C_k C_j \frac{1}{\alpha _k + \alpha
_j }} + \rho ^2\sum\limits_{k,j = 1}^{n_w } {D_k D_j \frac{1}{\beta _k +
\beta _j }} } ,
\]

where $\alpha _{i}$, $\beta _{i}$, $C_{i}$, $D_{i}$, $N$, $\rho $ are the
real model parameters; $n_u = n_w = 3$. The form of asymptotics in the limit
$r \to 0$ was assumed as: $u(r) \to r^2;w(r) \to r^3.$ The set of parameters
has to meet conditions

\[
\sum\limits_k {C_k = 0} ;\sum\limits_k {C_k \alpha _k = 0} ;\sum\limits_k
{D_k = 0} ;\sum\limits_k {D_k \frac{1}{\beta _k^2 } = 0} .
\]

In the limit $r \to \infty $ the deuteron wave functions must have such
known asymptotic form

\begin{equation}
\label{eq10}
\begin{array}{l}
 u(r) \to e^{ - \alpha r}; \\
 w(r) \to e^{ - \alpha r}\left( {1 + \frac{3}{\alpha r} + \frac{3}{(\alpha
r)^2}} \right), \\
 \end{array}
\end{equation}

where $\alpha = \sqrt {M\varepsilon } / \hbar = 0.2316$fm$^{ - 1}$;
$\varepsilon $=2.2245MeV is the deuteron binding energy. Then after the
application of the condition of equations (\ref{eq10}) to the deuteron wave
functions in forms (\ref{eq9}) leads to the relations for model parameters $\alpha
_1 = \beta _1 = \alpha $ and .

The charge and quadrupole deuteron form factors and the structure
function are defined by values of parameters of model. By using
this wave function has calculated the differential cross section
of the elastic deuteron-nucleus scattering \cite{Berezhnoy2005}.

The analytical form of DWF and it asymptotics, parameters of which
are determined by the power of asymptotic decrease of deuteron
form factors, particularly, the prediction of QCD, is obtained as
\cite{Krutov2004}:

\[
u(r) = \frac{32}{5\sqrt \pi \Gamma \left( {\frac{1}{4}}
\right)}\sum\limits_j {C_j m_j \left( {\frac{r}{2m_j }} \right)^{7 /
4}K_{\frac{3}{4}} (rm_j )} ,
\]

\[
\begin{array}{l}
 u(r \to 0) = r\frac{8\Gamma \left( {\frac{3}{4}} \right)}{5\sqrt \pi \Gamma
\left( {\frac{1}{4}} \right)}\sum\limits_j {\frac{C_j }{m_j^{3 / 2} }} ; \\
 u(r \to \infty ) = r^{5 / 4}\frac{32}{5\Gamma \left( {\frac{1}{4}}
\right)2^{9 / 4}}\sum\limits_j {\frac{C_j }{m_j^{5 / 4} }} e^{ - rm_j }, \\
 \end{array}
\]

or \cite{Krutov2006}:

\[
u(r) = \frac{128}{231\sqrt \pi \Gamma \left( {\frac{3}{4}}
\right)}\sum\limits_j {C_j m_j \left( {\frac{r}{2m_j }} \right)^{13 /
4}K_{\frac{9}{4}} (rm_j )} ,
\]

\[
\begin{array}{l}
 u(r \to 0) = r\frac{10\Gamma \left( {\frac{1}{4}} \right)}{231\sqrt \pi
\Gamma \left( {\frac{3}{4}} \right)}\sum\limits_j {\frac{C_j }{m_j^{9 / 2}
}} ; \\
 u(r \to \infty ) = r^{11 / 4}\frac{2^{13 / 4}}{231\Gamma \left(
{\frac{3}{4}} \right)}\frac{C_1 }{m_1^{11 / 4} }e^{ - \alpha r}, \\
 \end{array}
\]

where \textit{$\Gamma $(x)}, $K_{v}(x)$ are Euler and McDonald functions; $\alpha $=15/4.

If in addition to the conditions $\sum\limits_{j = 1}^N {C_j } =
0$ for the S- wave function the condition is imposed
\cite{Krutov2008} $\sum\limits_{j = 1}^N {C_j m_j^2 } = 0$, then
in the vicinity of zero the wave function has the following form

\[
u_0 (r) = r + ar^3;u_0^{ / / } (0) = 0.
\]

In Ref. \cite{Gamzova2009} research is conducted for pion
electromagnetic structure without asymptotic decomposition. It was
used the following wave function of Coulomb interaction at small
distances and linear confinement

\[
u(r) = N_T \exp \left( { - ar^{3/2} - \beta r} \right),
\]

where $\alpha = \frac{2}{3}\sqrt {2aM} $; $\beta = bM$; $a$ and $b$ are parameters
of linear and Coulomb parts of potential respectively.

The paper \cite{platonova2010} contains description of
spin-dependent observables in elastic proton-deuteron scattering
on the basis of a generalized diffraction model. This would have
parameterization for DWFs in coordinate space. To parameterize the
DWFs under consideration are employed used the sum of Gaussian
functions with account the behavior of the wave functions at $r$=0

\[
\begin{array}{l}
 u(r) = r\sum\limits_{j = 1}^m {C_{0j} \exp ( - A_{0j} r^2)} , \\
 w(r) = r^3\sum\limits_{j = 1}^m {C_{2j} \exp ( - A_{2j} r^2)} , \\
 \end{array}
\]

where $m$=5. The functions fitted on the basis of numerical values
for the CD-Bonn and dressed dibaryon model (DBM) functions in the
intervals 0-20fm with a step of 0.1fm.

In \cite{Babenko2011} are specified results of calculations for
the deuteron quadrupole momentum $Q$ by using experimental phase
shifts for partial-wave analysis of GWU (George Washington
University) \cite{Arndt2000} and Nijmegen \cite{Stoks1993}. Also
the deuteron parameters (deuteron quadrupole moment Q, the
deuteron asymptotic D/S and the deuteron asymptotic normalization
constant A$_{S})$ and correlation between them for the group
potentials is studied. This dependence is represent in the form  $Q/ \eta = a + bA_S^2 $,

where $a$=3.92464fm$^{2}$; $b$=8.71829fm$^{3}$.

Influence of the D- state component of DWF \cite{Adler1977} on the
application of the Trojan horse method it was shown in
\cite{Lamia2012}.

Parametrization formulas in a form according to \cite{Lacombe1981}
are applied approximations of DWF for potential charge-dependent
Bonn (CD-Bonn) \cite{Machleidt2001}, model FSS2 with the Coulomb
exchange kernel \cite{Fujiwara2001}, and calculated in three
different schemes (isospin basis and particle basis with Coulomb
off or Coulomb on) and fss2 baryon-baryon interaction
\cite{Fukukawa2015} at$ N$=11, and also for MT model
\cite{Krutov2007} when $N_{S}$=16; $N_{D}$=12.

Parametrization Dubovichenko \cite{Dubovichenko2004} is improved
in works \cite{Zhaba1, Zhaba2, Zhaba3, Zhaba4}. Minimization of
values $\chi ^{2}$ is carried out 10$^{ - 4}$. Using deuteron wave
functions in coordinate and space representations, are designed a
component of a tensor of sensitivity polarization of deuterons
$T_{20}$ \cite{Karmanov1981} polarization transmission $K_{0}$,
tensor analyzing power $A_{yy}$ and tensor-tensor transmission of
polarization $K_{y}$ \cite{Ladygin2002}. The obtained outcomes are
compared to the published experimental and theoretical outcomes

For deuteron wave function in configuration representation for
potential Argonne v18 are designed numerical coefficients of
analytical forms \cite{Zhaba5}

\[
\left\{ {\begin{array}{l}
 u(r) = \sum\limits_{i = 1}^{20} {A_i \exp ( - a_i r^3),} \\
 w(r) = r^2\sum\limits_{i = 1}^{20} {B_i \exp ( - b_i r^3).} \\
 \end{array}} \right.
\]

The coefficients of the four approximating dependencies for the
numerical values of DWFs for four realistic phenomenological
potentials Nijmegen group have been numerically calculated. The
analytical forms are chosen as the product of the power function
r$^{n}$ for the sum of exponential terms \cite{Zhaba6}:

\[
\left\{ {\begin{array}{l}
 u(r) = r\sum\limits_{i = 1}^N {A_i \exp ( - a_i r^2),} \\
 w(r) = r\sum\limits_{i = 1}^N {B_i \exp ( - b_i r^2),} \\
 \end{array}} \right.
\]

\[
\left\{ {\begin{array}{l}
 u(r) = r^2\sum\limits_{i = 1}^N {A_i \exp ( - a_i r^3),} \\
 w(r) = r^2\sum\limits_{i = 1}^N {B_i \exp ( - b_i r^3).} \\
 \end{array}} \right.
\]

The behavior of the value $\chi ^{2}$ depending on the number of
expansion terms $N_{i}$ has been studied. With the account of the
minimum values of $\chi ^{2}$ for these forms we have built DWFs
in the coordinate space, which do not contain superfluous knots.
The calculated parameters of the deuteron are in good agreement
with theoretical and experimental results. For DWFs in coordinate
and momentum space it is calculated such polarization
characteristics: the tensor polarization \cite{Garson1994} (values
$t_{20}(p)$, $t_{21}(p)$, $t_{22}(p))$ in the range of 0-7 pulse
fm$^{-1}$. The value of $t_{20}(p)$ for potentials Nijmegen group
in good agreement with literature results for other potential
nucleon-nucleon of models and with experimental data's. The
results of the deuteron tensor polarization $t_{ij}(p)$ give some
information about the electromagnetic structure of the deuteron.
And when known tensor analyzing power it is possible to calculate
the differential cross section of double scattering.

To solve the system of associated Schr\"{o}dinger equations that describe
the radial DWF $u$ and $w$

\[
\left\{ {\begin{array}{l}
 u'' - \alpha ^2u = f(r), \\
 w'' - \left( {\alpha ^2 + \frac{6}{r^2}} \right)w = g(r) \\
 \end{array}} \right.
\]

parameterizations were proposed back in 1955 \cite{Cap1955}:

\[
\left\{ {\begin{array}{l}
 f(r) = \sum\limits_{n = 0}^\infty {c_n \psi _{1n} (r)} , \\
 g(r) = \sum\limits_{n = 0}^\infty {d_n \psi _{1n} (r)} . \\
 \end{array}} \right.
\]

They can be generalized for the DWF approximation as such
analytical forms through Laguerre functions \cite{Cap1955}:

\[
\left\{ {\begin{array}{l}
 u(r) = \sum\limits_{n = 0}^{11} {A_n \psi _{3n} (r),} \\
 w(r) = \sum\limits_{n = 0}^{11} {B_n \psi _{3n} (r),} \\
 \end{array}} \right.
\]

where $\psi _{3n} (r)$ - Laguerre functions ($n$=0,1,2,3,\ldots ):

\[
\psi _{3n} (r) = \frac{2\alpha \sqrt {2\alpha } }{n!\sqrt {(n + 1)(n + 2)}
}\frac{\exp (\alpha r)}{r}\frac{d^n}{dr^n}\left( {r^{n + 2}\exp ( - 2\alpha
r)} \right),
\]

\[
\psi _{30} = \sqrt \alpha \exp ( - \alpha r)\left( {2\alpha r} \right),
\]

\[
\psi _{31} = 2\sqrt {\frac{\alpha }{3}} \exp ( - \alpha r)\left( {3\alpha r
- 2\alpha ^2r^2} \right),
\]

\[
\psi _{32} = 2\sqrt {\frac{2\alpha }{3}} \exp ( - \alpha r)\left( {3\alpha r
- 4\alpha ^2r^2 + \alpha ^3r^3} \right),
\]

\[
\psi _{33} = 2\sqrt {10\alpha } \exp ( - \alpha r)\left( {\alpha r - 2\alpha
^2r^2 + \alpha ^3r^3 - \frac{2}{15}\alpha ^4r^4} \right),
\]

\[
\psi _{34} = \sqrt {\frac{5\alpha }{3}} \exp ( - \alpha r)\left( {6\alpha r
- 16\alpha ^2r^2 + 12\alpha ^3r^3 - \frac{16}{5}\alpha ^4r^4 +
\frac{4}{15}\alpha ^5r^5} \right),
\]

\[
\psi _{35} = 2\sqrt {\frac{7\alpha }{3}} \exp ( - \alpha r)\left( {3\alpha r
- 10\alpha ^2r^2 + 10\alpha ^3r^3 - 4\alpha ^4r^4 + \frac{2}{3}\alpha ^5r^5
- \frac{4}{105}\alpha ^6r^6} \right).
\]

The coefficients of analytical forms through Laguerre functions
for the deuteron wave function in coordinate space for NijmI,
NijmII, Nijm93, Reid93 and Argonne v18 potentials have been
numerically calculated in \cite{Zhaba7}. Near the beginning of
coordinates there are some small oscillations for DWFs, but
despite of it designed static parameters well coincide with
original values.

Parameterizations \cite{Cap1955} and \cite{Dubovichenko20002} can
be generalized for the DWF approximation as such analytical forms:

\begin{equation}
\label{eq11}
\left\{ {\begin{array}{l}
 u(r) = r^A\sum\limits_{i = 1}^N {A_i \exp ( - a_i r^3),} \\
 w(r) = r^B\sum\limits_{i = 1}^N {B_i \exp ( - b_i r^3).} \\
 \end{array}} \right.
\end{equation}

Given $N$=11, search for an index of function of a degree $r^{n}$
has been carried out, appearing as a factor before the sums of
exponential terms of the analytical form (\ref{eq11}). Best values
appeared to be $n$=1.47 and $n$=1.01 for $u(r)$ and $w(r)$
accordingly. Hence, the factors before the sums in (\ref{eq11})
can be chosen as $r^{3 / 2}$ and $r^{1}$ \cite{Zhaba8}:

\begin{equation}
\label{eq12}
\left\{ {\begin{array}{l}
 u(r) = r^{3 / 2}\sum\limits_{i = 1}^N {A_i \exp ( - a_i r^3),} \\
 w(r) = r\sum\limits_{i = 1}^N {B_i \exp ( - b_i r^3).} \\
 \end{array}} \right.
\end{equation}

Despite cumbersome and time-consuming calculations and
minimizations of \textit{$\chi $}$^{2}$ (to the value smaller than
10$^{-7})$, it was necessary to approximate numerical values of
DWF, the arrays of numbers of which made up 839х4 values in an
interval $r$=0-25~fm for potentials NijmI, NijmII, Nijm93 and
Reid93 \cite{stoks1994}, and 1500х2 values in an interval$
r$=0-15~fm for potential Argonne v18 \cite{Wiringa1995}.

The accuracy of parametrization (\ref{eq12}) is characterized by:

\[
\chi ^2 = \frac{1}{n - p}\sum\limits_{i = 1}^N {\left( {y_i - f(x_i ;a_1
,a_2 ,...,a_p )} \right)^2} ,
\]

where $n$ - the number of points of the array $y_{i}$ of the numerical values of
DWF in the coordinate space; $f$ - approximating function of $u$ (or $w)$ according
to the formulas (\ref{eq2}); $a_{1}$,$a_{2}$,\ldots ,$a_{p}$ - parameters; $p$ - the
number of parameters (coefficients in the sums of formulas (\ref{eq12})). Hence,
\textit{$\chi $}$^{2}$ is determined not only by the shape of the approximating function
$f$, but also by the number of the selected parameters.

The approximation can be made on the whole interval, or divided into a few
distinct sites: around the origin in the maximum and descending function.
But this complicates further generalization for the form of the wave
function.

Coefficients and DWFS (\ref{eq12}) for NijmI, NijmII, Nijm93,
Reid93 and Argonne v18 potentials it is resulted in works
\cite{Zhaba9, Zhaba10}. A detailed comparison of the obtained
values of $t_{20}(p)$ (the scattering angle \textit{$\theta
$=}70$^{0})$ for these potentials with the up-to-date experimental
data of JLAB t20 \cite{Abbott2000a, Abbott2000b} and BLAST
\cite{Garson1994, Zhang2011} collaborations. There is a good
agreement is for the momentas $p$=1-4~fm~$^{ - 1}$.

If we consider normalization $\int {(u^2 + w^2)dr} = 1$ for DWFs (\ref{eq11}), we
can write this condition using the corresponding coefficients as

\[
\sum\limits_{i = 1}^N {\left( {\frac{2^{2 / 3}\Gamma \left[ {\frac{4}{3}}
\right]A_i^2 }{12a_i^{4 / 3} } + \frac{B_i^2 }{6b_i }} \right)} = 1.
\]

In this paper it has been used parameterization (\ref{eq12}) and
it is made minimization of quantity of the designed coefficients.
Dependence \textit{$\chi $}$^{2}$ from the number of expansion
terms $N$ is resulted in Tables 5 and 6 separately for functions
$u(r)$ and $w(r)$. At increase for value $N$ reduction of size
\textit{$\chi $}$^{2}$ for \emph{u(r)} (potential Reid93) is
precisely shown in Fig. 3. The coefficients of new analytical
forms for DWF in coordinate space for NijmI, NijmII, Nijm93,
Reid93 and Argonne v18 potentials have been numerically calculated
(Tables 7-11). The obtained wave functions (Fig. 4 and 5) do not
contain any superfluous knots.

Based on the known DWFs (\ref{eq12}) and them coefficients (Tables
7-11) one can calculate the deuteron properties (Table 12):
deuteron radius $r_{m}$, the quadrupole moment $Q_{d}$,the $D$-
state probability $P_{D}$ and the magnetic moment \textit{$\mu
$}$_{d}$. They are in good agreement with the theoretical (Table
1) and experimental (Table 2) data.

\pdfximage width 135mm {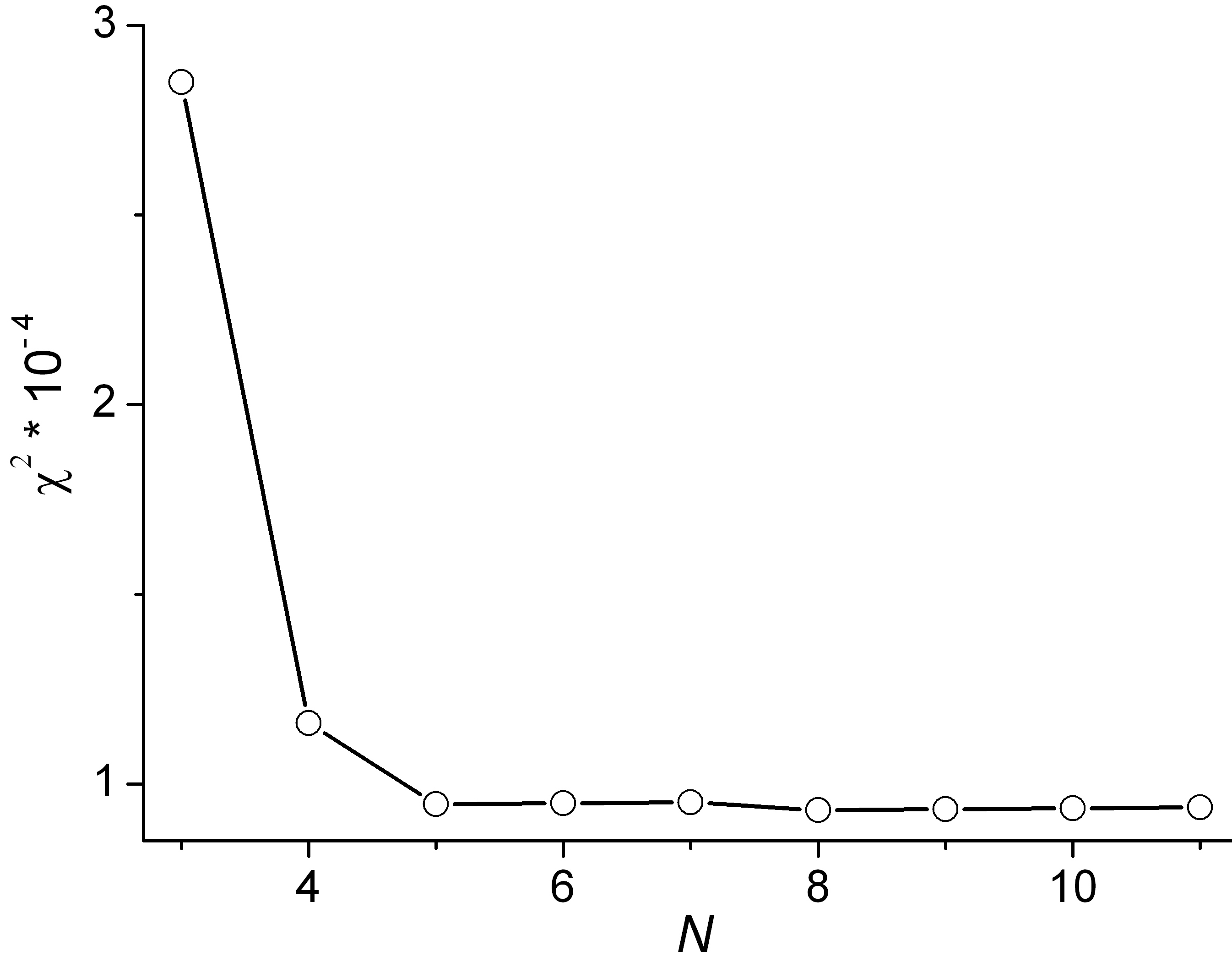}\pdfrefximage\pdflastximage

Fig. 3. \textit{$\chi $}$^{2}$ for \emph{u(r)} (potential Reid93)

\pdfximage width 135mm {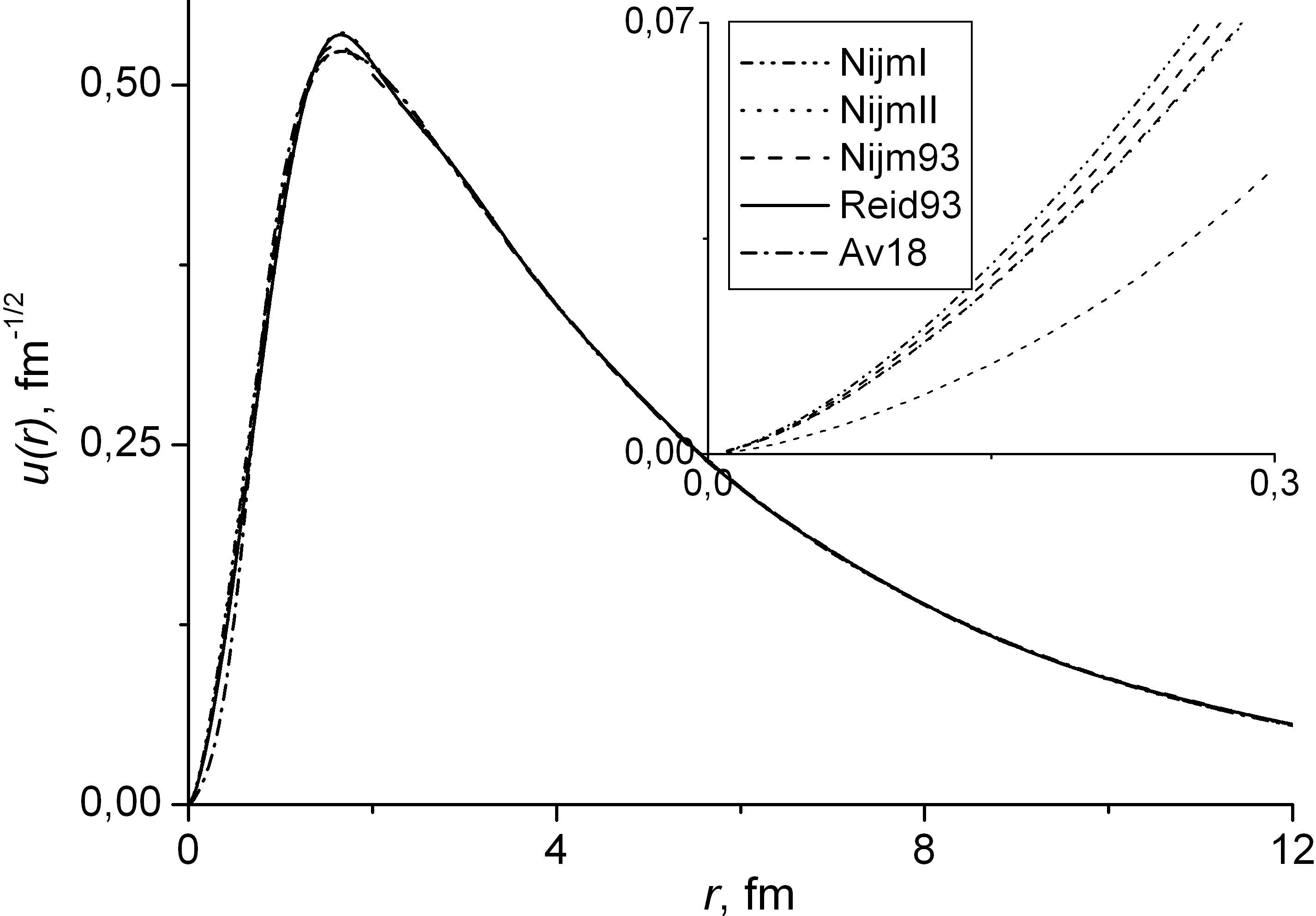}\pdfrefximage\pdflastximage

Fig. 4. Deuteron wave function \emph{u(r)}

\pdfximage width 135mm {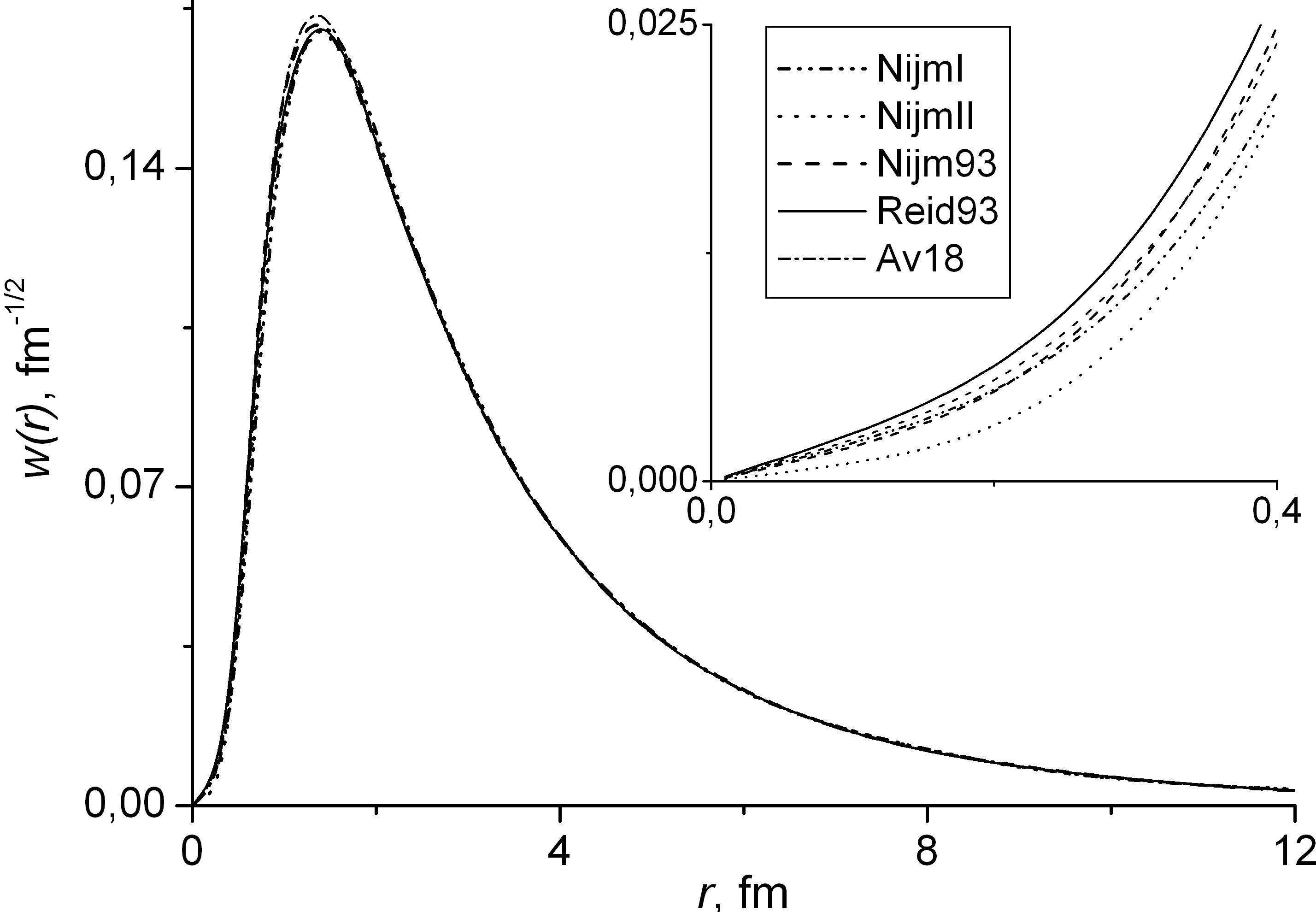}\pdfrefximage\pdflastximage

Fig. 5. Deuteron wave function \emph{w(r)}

\begin{center}
Table 5. Values \textit{$\chi $}$^{2}$ for $u(r)$
\end{center}

\begin{table}[htbp]
\begin{tabular}
{|p{32pt}|p{56pt}|p{63pt}|p{58pt}|p{58pt}|p{58pt}|}
\hline
$N $&
NijmI &
NijmII &
Nijm93 &
Reid93 &
Av18 \\
\hline
3&
2.61E-04&
3.78E-04&
2.64E-04&
2.85E-04&
2.24E-05 \\
\hline
4&
1.13E-04&
2.13E-04&
6.66E-05&
1.16E-04&
3.26E-06 \\
\hline
5&
1.76E-05&
1.92E-04&
3.92E-05&
9.47E-05&
6.45E-07 \\
\hline
6&
1.18E-05&
1.92E-04&
3.93E-05&
9.49E-05&
4.27E-07 \\
\hline
7&
1.08E-05&
1.93E-04&
3.61E-05&
9.51E-05&
4.24E-07 \\
\hline
8&
1.27E-06&
1.91E-04&
3.62E-05&
9.31E-05&
4.25E-07 \\
\hline
9&
1.30E-06&
1.94E-04&
3.63E-05&
9.33E-05&
4.10E-07 \\
\hline
10&
8.03E-06&
1.92E-04&
3.64E-05&
9.36E-05&
4.03E-07 \\
\hline
11&
7.01E-06&
1.93E-04&
3.64E-05&
9.38E-05&
4.04E-07 \\
\hline
\end{tabular}
\label{tab5}
\end{table}

\begin{center}
Table 6. Values \textit{$\chi $}$^{2}$ for $w(r)$
\end{center}

\begin{table}[htbp]
\begin{tabular}
{|p{32pt}|p{56pt}|p{63pt}|p{58pt}|p{58pt}|p{58pt}|}
\hline
$N $&
NijmI &
NijmII &
Nijm93 &
Reid93 &
Av18 \\
\hline
3&
2.56E-05&
2.72E-05&
3.10E-05&
2.82E-05&
4.46E-06 \\
\hline
4&
2.81E-06&
2.73E-06&
3.45E-06&
3.20E-06&
3.58E-06 \\
\hline
5&
8.46E-07&
4.94E-07&
7.50E-07&
1.01E-06&
3.58E-06 \\
\hline
6&
6.52E-07&
2.77E-07&
4.49E-07&
7.96E-07&
3.58E-06 \\
\hline
7&
6.53E-07&
2.68E-07&
4.29E-07&
7.97E-07&
4.30E-07 \\
\hline
8&
6.52E-07&
2.59E-07&
4.20E-07&
7.99E-07&
4.31E-07 \\
\hline
9&
6.56E-07&
2.58E-07&
4.21E-07&
8.00E-07&
4.32E-07 \\
\hline
10&
6.58E-07&
2.60E-07&
4.19E-07&
7.75E-07&
4.32E-07 \\
\hline
11&
6.84E-07&
2.61E-07&
4.23E-07&
7.86E-07&
4.31E-07 \\
\hline
\end{tabular}
\label{tab6}
\end{table}

\begin{center}
Table 7. Coefficients $A_{i}$, $a_{i}$, $B_{i}$, $b_{i}$ (NijmI)
\end{center}

\begin{table}[htbp]
\begin{tabular}
{|p{16pt}|p{85pt}|p{85pt}|p{89pt}|p{85pt}|}
\hline
$i$&
$A_{i}$&
$a_{i}$&
$B_{i}$&
$b_{i} \quad  $ \\
\hline
1&
0.00065826309&
0.00023679107&
-0.15255613149&
3.97342104241 \\
\hline
2&
0.04005454537&
0.01185654423&
0.00081779640&
0.00060963929 \\
\hline
3&
0.01595367314&
0.00350795838&
0.04767548848&
0.22931527097 \\
\hline
4&
0.20572009408&
0.42747031442&
0.01223007130&
0.22931527246 \\
\hline
5&
0.08430196890&
0.03911783132&
0.05400988310&
0.06026407442 \\
\hline
6&
0.15463267892&
0.12780158776&
0.00599092790&
0.00361063044 \\
\hline
7&
0.00466297503&
0.00098105003&
0.03019488106&
0.22931528067 \\
\hline
8&
0.02431014420&
4.49214201019&
0.02219993378&
0.01578590745 \\
\hline
\end{tabular}
\label{tab7}
\end{table}

\begin{center}
Table 8. Coefficients $A_{i}$, $a_{i}$, $B_{i}$, $b_{i}$ (NijmII)
\end{center}

\begin{table}[htbp]
\begin{tabular}
{|p{16pt}|p{85pt}|p{85pt}|p{89pt}|p{85pt}|}
\hline
$i$&
$A_{i}$&
$a_{i}$&
$B_{i}$&
$b_{i} \quad  $ \\
\hline
1&
0.00085796423&
0.00026671465&
-0.16642454661&
4.70854300660 \\
\hline
2&
0.05148122494&
0.01601293924&
0.00093664870&
0.00066353652 \\
\hline
3&
0.00615093450&
0.00118853942&
0.01905670791&
0.25657778517 \\
\hline
4&
0.11879048059&
0.05800341799&
0.00671655150&
0.00400272658 \\
\hline
5&
0.09493719253&
0.25110911883&
0.05706931441&
0.06798490110 \\
\hline
6&
0.09222467415&
0.25162731987&
0.01956153732&
0.25657778499 \\
\hline
7&
0.08013045048&
0.25146410450&
0.02445879052&
0.01769592277 \\
\hline
8&
0.02054925493&
0.00451468381&
0.01976865474&
0.25657778490 \\
\hline
9&
&
&
0.02834209930&
0.25657779090 \\
\hline
\end{tabular}
\label{tab8}
\end{table}

\begin{center}
Table 9. Coefficients $A_{i}$, $a_{i}$, $B_{i}$, $b_{i}$ (Nijm93)
\end{center}

\begin{table}[htbp]
\begin{tabular}
{|p{22pt}|p{85pt}|p{85pt}|p{89pt}|p{85pt}|}
\hline
$i$&
$A_{i}$&
$a_{i}$&
$B_{i}$&
$b_{i} \quad  $ \\
\hline
1&
0.00098878586&
0.00028491272&
-0.16660171842&
5.01389130303 \\
\hline
2&
0.13571076271&
0.07486587795&
0.00042682696&
0.00042106442 \\
\hline
3&
0.12400573946&
0.32247840206&
0.08172183510&
0.26642919122 \\
\hline
4&
0.14274952979&
0.32247840078&
-0.13944102365&
0.19870727465 \\
\hline
5&
0.06212705650&
0.02002538042&
0.09969921461&
0.11841769371 \\
\hline
6&
0.00719613810&
0.00132675713&
0.05422237632&
0.26560362260 \\
\hline
7&
0.02453978059&
0.00531708637&
0.03879995997&
0.26603771790 \\
\hline
8&
&
&
0.00329896732&
0.00224533303 \\
\hline
9&
&
&
0.01264941157&
0.00905020138 \\
\hline
10&
&
&
0.03368130195&
0.03188653927 \\
\hline
\end{tabular}
\label{tab9}
\end{table}

\begin{center}
Table 10. Coefficients $A_{i}$, $a_{i}$, $B_{i}$, $b_{i}$ (Reid93)
\end{center}

\begin{table}[htbp]
\begin{tabular}
{|p{22pt}|p{85pt}|p{85pt}|p{89pt}|p{85pt}|}
\hline
$i$&
$A_{i}$&
$a_{i}$&
$B_{i}$&
$b_{i} \quad  $ \\
\hline
1&
0.00085859852&
0.00026661728&
-0.15112264940&
5.27032534000 \\
\hline
2&
0.02106750054&
0.00457697258&
0.00032452355&
0.00036454826 \\
\hline
3&
0.00620290953&
0.00119165642&
-0.11845656993&
0.15690222137 \\
\hline
4&
0.11117085753&
0.25958967300&
0.05880984638&
0.21866245485 \\
\hline
5&
0.12121298131&
0.06065662499&
0.05571966351&
0.21983692771 \\
\hline
6&
0.09712838383&
0.25946221474&
0.02824394399&
0.02520241662 \\
\hline
7&
0.05353538391&
0.01656444102&
0.04817589390&
0.21942325873 \\
\hline
8&
0.05686488199&
0.25952108094&
0.00253775036&
0.00184332786 \\
\hline
9&
&
&
0.09143245521&
0.09410956021 \\
\hline
10&
&
&
0.00999601651&
0.00723824613 \\
\hline
\end{tabular}
\label{tab10}
\end{table}

\begin{center}
Table 11. Coefficients $A_{i}$, $a_{i}$, $B_{i}$, $b_{i}$ (Av18)
\end{center}

\begin{table}[htbp]
\begin{tabular}
{|p{22pt}|p{89pt}|p{85pt}|p{89pt}|p{85pt}|}
\hline
$i$&
$A_{i}$&
$a_{i}$&
$B_{i}$&
$b_{i} \quad  $ \\
\hline
1&
-2.31737065809&
0.35274596179&
-0.16442140495&
4.27551981801 \\
\hline
2&
0.02659110800&
0.00732058085&
0.02868983216&
0.02949360502 \\
\hline
3&
-0.28877337807&
5.38000986025&
0.00074392415&
0.00062004296 \\
\hline
4&
0.99922786191&
0.36067988379&
0.05707754763&
0.09089766773 \\
\hline
5&
0.11094070754&
0.06921749879&
0.01167238661&
0.00953758791 \\
\hline
6&
0.01077521733&
0.00225041184&
0.00370587438&
0.00281824630 \\
\hline
7&
0.00274077964&
0.00052965934&
0.08509685731&
0.30274669392 \\
\hline
8&
0.65409540449&
0.26723780768&
&
 \\
\hline
9&
0.99185931871&
0.41270669565&
&
 \\
\hline
10&
0.05693893648&
0.02247865903&
&
 \\
\hline
\end{tabular}
\label{tab11}
\end{table}

\begin{center}
Table 12. Deuteron properties
\end{center}

\begin{table}\begin{tabular}{|p{68pt}|p{49pt}|p{55pt}|p{49pt}|p{55pt}|}
\hline Potential& $r_{m}$ (fm)& $Q_{d}$ (fm$^{2})$& $P_{D}$
({\%})&
\textit{$\mu $}$_{d}$  \\
\hline NijmI& 1.96616& 0.271372& 5.65618&
0.847577 \\
\hline NijmII& 1.96711& 0.270014& 5.62972&
0.847727 \\
\hline Nijm93& 1.96543& 0.270362& 5.74951&
0.847045 \\
\hline Reid93& 1.96819& 0.270162& 5.69023&
0.847383 \\
\hline Argonne v18& 1.95471& 0.268201& 5.75946&
0.846988 \\
\hline
\end{tabular}
\label{tab12}
\end{table}

\textbf{8. Conclusions}

Static properties of the deuteron ($E_{d}$ , $r_{m}$, $Q_{d}$ ,
$P_{D}$ , $\eta$ , $A_{S})$, obtained by DWFs for potential
models, have been chronologically systematized. The presence or
absence of knots near the origin of coordinates for the radial DWF
have been shown. The forms, methods of obtaining and asymptotic
behaviors of analytic forms for DWFs in the coordinate space have been analyzed.

Parameterization in the form of (\ref{eq12}) has been used and the
number of expansion coefficients has been minimized. Dependence of
$\chi^{2}$ on the number of expansion terms N parameterization
(\ref{eq12}) is shown separately for the functions \emph{u(r)} and
\emph{w(r}). The optimum is \emph{N}=7-10. The resulting wave
functions do not contain any extra knots. Calculations have been
done for realistic phenomenological potentials NijmI, NijmII,
Nijm93, Reid93 and Argonne v18. What is more, analytical forms of
DWF by such authors as Сertov, Mathelitsch, Moravcsik and
Machleidt have been "improved".

The resulting DWFs for the group of potential models can be
applied to calculate polarization characteristics of the deuteron
(tensor polarization $t20$, sensitivity tensor component to
polarization of deuterons $T_{20}$, polarization transmission
$K_{0}$ and tensor analyzing power $A_{yy}$, etc.
\cite{Ladygin2002}). The results will allow studying the deuteron
electromagnetic structure, its form-factors and differential cross
section of double scattering in more detail in future.

\end{document}